\documentclass[twocolumn,amssymb,aps,prc,showpacs,floatfix]{revtex4-1}

\usepackage{float}
\usepackage{graphicx}
\usepackage{amsmath}

\begin{document}

\title{Neutrino diffusive transport in hot quark matter: a detailed analysis}
\author{Gustavo C. Colvero and  Germ\'an Lugones}
\affiliation{Universidade Federal do ABC, Rua Santa Ad\'elia 166, Santo Andr\'e, SP, 09210-170, 
Brazil.}
\email{gustavo.colvero@gmail.com} 
\email{german.lugones@ufabc.edu.br} 

\received{17 February 2014}
\published{......}

%%%%%%%%%%%%%%%%%%%%%%%%%%%%%%%%%%%%%%%%%%%%%%%%%%%%%%%%%%%%%%%%%%%%%%%%%%%%%%%%%%%%%%%%%%%%%%%%%%%% ABSTRACT
% ABSTRACT                                                                                         %
%%%%%%%%%%%%%%%%%%%%%%%%%%%%%%%%%%%%%%%%%%%%%%%%%%%%%%%%%%%%%%%%%%%%%%%%%%%%%%%%%%%%%%%%%%%%%%%%%%%%

\begin{abstract}
We perform an extensive analysis of neutrino diffusion in quark matter within the 
MIT bag model at arbitrary temperature and degeneracy. We examine in detail the contribution of each 
of the relevant weak interaction processes to the total neutrino opacity and evaluate the effect 
of the strange quark mass, the bag constant, and the QCD perturbative corrections to the MIT bag model. 
We also investigate the anisotropic contribution to the neutrino distribution function in 
scatterings, the mean energy transfer and the mean scattering angle. The density and temperature 
dependence of the diffusion coefficients $D_n$ that govern the cooling and deleptonization of a 
compact star is shown in detail. Finally, our numerical results for the neutrino mean free paths 
are compared against known analytic approximations. 
We conclude that neutrino scattering constitutes a significant portion of the total neutrino opacity in 
leptonized quark matter and neutrino-quark scattering is, in general, very similar to 
neutrino-electron scattering with respect to both mean energy transfer per scattering and mean 
scattering angle.

\end{abstract}

\pacs{97.60.Jd, 21.65.Qr, 13.15.+g}

\maketitle

%%%%%%%%%%%%%%%%%%%%%%%%%%%%%%%%%%%%%%%%%%%%%%%%%%%%%%%%%%%%%%%%%%%%%%%%%%%%%%%%%%%%%%%%%%%%%%%%%%%% SEC
% INTRO                                                                                            % INTRO
%%%%%%%%%%%%%%%%%%%%%%%%%%%%%%%%%%%%%%%%%%%%%%%%%%%%%%%%%%%%%%%%%%%%%%%%%%%%%%%%%%%%%%%%%%%%%%%%%%%% 

\section{Introduction}
Compact stellar objects containing deconfined quark matter have been envisaged 
since long ago \cite{Bodmer1971,Terazawa1979,Witten1984,Alcock1986,Haensel1986}. According to 
theoretical studies such objects could be formed if the density inside a purely hadronic star goes 
beyond a critical density  \cite{critical_density}.  This may happen due to accretion onto a cold 
hadronic star in a binary system, as a consequence of cooling,  deleptonization and fallback 
accretion during the protoneutron star phase of a just born compact star, due to spin down of a 
fast rotating star  or by more exotic mechanisms such as strangelet contamination 
\cite{Glendenning_book,Weber_book,Haensel_book}.  The conversion of the star presumably begins 
through the formation of a small quark matter seed that grows at the expenses of the gravitational 
energy extracted from the contraction of the object and/or through strongly exothermic combustion 
processes \cite{Lugones2002,Keranen2005,Niebergal2010,Fischer2011,Pagliara2013}. All these 
scenarios lead to the formation of hot and neutrino rich quark matter occupying the whole 
compact object (strange star) or its core (hybrid star).

A full comprehension of the above mechanisms requires a detailed knowledge of 
the neutrino transport in quark matter. In fact, 
neutrino interactions in neutron star matter have been studied in detail for a long time now. Most 
of the early contributions were given in the form of approximate treatments focused on a particular 
state of neutrino degeneracy: neutrinos were considered to be either nondegenerate or highly 
degenerate. From these works, those which are particularly important for the present analysis have 
been summarized by Iwamoto \cite{Iwamoto1982}, with emphasis in neutrino interactions in quark 
matter. 

Extensive numeric analyses for arbitrary temperatures, densities and neutrino level of degeneracy, 
however, are not so abundant. In this case, we may cite the works of Reddy \& Prakash 
\cite{Reddy1997} and Reddy, Prakash \& Lattimer \cite{Reddy1998}, where neutrino interactions in 
hadronic neutron star matter with the possible presence of hyperons have been studied.

Works focused on neutrino interactions in quark matter at arbitrary conditions are even rarer and 
the reference for this topic is usually the paper of Steiner, Prakash \& Lattimer 
\cite{Steiner2001}. In that work, the authors analyse the neutrino diffusion coefficients in the 
interior of a hybrid star containing a mixed quark-hadron phase and quark matter was described
using the MIT bag model with a fixed bag constant $B=200$MeV/fm$^3$.

In the present work, we intend to fill the gap in the literature represented by the lack of studies 
exploring general aspects of neutrino interactions and diffusive neutrino transport in quark matter 
in general conditions. We describe quark matter using the MIT bag model and analyze the influence of its three 
possible free parameters $-$ the bag constant $B$, the mass of the strange quark $m_s$ and the strong 
coupling constant $\alpha_c$ $-$ on the neutrino mean free paths and energy-averaged diffusion
coefficients. We pay particular attention to the relative contributions of absorption 
and scattering to the total neutrino opacity and how neutrino-quark scattering compares to 
neutrino-electron scattering. Having time-consuming simulations in mind, we explore different 
expressions for the neutrino scattering opacity and the possible penalties in precision associated 
to more simplified treatments. We also explore here the actual range of validity of the known 
analytic approximate forms for the neutrino mean free paths.

%%%%%%%%%%%%%%%%%%%%%%%%%%%%%%%%%%%%%%%%%%%%%%%%%%%%%%%%%%%%%%%%%%%%%%%%%%%%%%%%%%%%%%%%%%%%%%%%%%%% SEC
% BTE DIFFUSION                                                                                    % BTE DIFFUSION
%%%%%%%%%%%%%%%%%%%%%%%%%%%%%%%%%%%%%%%%%%%%%%%%%%%%%%%%%%%%%%%%%%%%%%%%%%%%%%%%%%%%%%%%%%%%%%%%%%%% 

\section{Boltzmann transport in the diffusive regime}\label{sec:bte}
Neutrino transport is described in terms of the Boltzmann transport equation for massless particles, 
which dictates the kinetic evolution of its invariant distribution function $f=f(x,p)$:
\begin{align}\label{eq:BTE_general}
  p^\alpha\frac{Df}{d x^\alpha}=\left(\frac{df}{d\tau}\right)_{\mbox{\scriptsize{coll.}}},
\end{align}
\noindent where, in the general relativistic case, $D/dx^\alpha$ denotes the operator 
$\partial /\partial x^\alpha-\Gamma^{\beta}_{\alpha \gamma}p^\gamma \partial/\partial p^\beta$, with 
$\Gamma^{\beta}_{\alpha \gamma}$ being the Christoffel symbols. The right-hand side of 
Eq. (\ref{eq:BTE_general}) designates the changes in $f$ due to particle interaction processes 
(``collisions") and it can be separated into different contributions
\begin{align}\label{eq:ColInt_B}
  \left(\frac{df}{d\tau}\right)_{\mbox{\scriptsize{coll.}}}=B_{\mbox{\scriptsize{AE}}}[f]
		+B_{\mbox{\scriptsize{S}}}[f]+\cdots,
\end{align}
\noindent where AE stands for absorption/emission of neutrinos and S designates all forms
of scattering. Here, we will focus only on these two kind of processes, since those are the dominant 
ones on neutron star matter at the typical values of temperature and lepton degeneracy we are 
interested in. For processes of the form $\nu+2\rightarrow 3+4$, each contribution to
Eq. (\ref{eq:ColInt_B}) can be explicitly written in the generic ``collision integral"
\begin{align}\label{eq:CollisionInt}
  &B\left[f\right]=-\int\frac{d^3p_2}{(2\pi)^3}\int\frac{d^3p_3}{(2\pi)^3}
	\int\frac{d^3p_4}{(2\pi)^3}\Big\{\nonumber\\
	&\times g_\nu g_2f(E_\nu)f_2(E_2)[1-f_3(E_3)][1-f_4(E_4)]W_{fi}\nonumber\\
	&-g_3g_4[1-f(E_\nu)][1-f_2(E_2)]f_3(E_3)f_4(E_4)W_{if}\Big\}
\end{align}
\noindent where $f_i$ denotes the distribution function for particle species $i$, $f$ is the 
complete neutrino distribution function and the $g_i$ denote the phase-space spin degeneracy for 
each particle. Here, $W_{fi}$ represents the transition rate between the initial and the final 
states for the given process. We can define the neutrino emissivity $j_a$ and the absorptivity 
$1/ \lambda_a$ through 
\begin{align}\label{eq:BAE}
  B_{\mbox{\scriptsize{AE}}}[f]&=\left[1-f(E_\nu)\right]j_a-\frac{f(E_\nu)}{\lambda_a},
\end{align}	
\noindent and the scattering contribution to the collision integral can be written as
\begin{align}\label{eq:BScatt}
	B_{\mbox{\scriptsize{S}}}[f]=&\left[1-f(E_\nu)\right]\int\frac{d^3p_4}{(2\pi)^3}f(E_\nu^\prime)	
  R^{in}\nonumber\\
	&-f(E_\nu)\int\frac{d^3p_4}{(2\pi)^3}\left[1-f(E_\nu^\prime)\right]R^{out},
\end{align}
\noindent where $R^{in/out}(E_\nu,E_\nu^\prime,\cos\theta)$ 
are the scattering kernels, $\theta$ being the scattering angle. When  the pairs of reactions of
emission/absorption and scattering in/out are balanced, the transition rates satisfy the reciprocity 
relations $g_\nu g_2 W_{if}=g_3g_4 W_{if}$ and we have the relations of detailed balance
\begin{align}\label{eq:DetBalance}
  \frac{1}{\lambda_a}=e^{(E_\nu-\mu_\nu)/T}j_a,		\quad R^{in}=e^{(E_\nu^\prime-E_\nu)/T}R^{out},
\end{align}
\noindent \textit{i.e.}, Eqs. (\ref{eq:BAE}) and (\ref{eq:BScatt}) have only one independent 
contribution each.

The \textit{diffusive approximation} consists of assuming that matter is at thermodynamic 
equilibrium and neutrinos are only slightly out of equilibrium. Since the equilibrium states of our 
MIT bag model are those of a gas non-interacting fermions, we automatically set each $f_i$ in 
Eq. (\ref{eq:CollisionInt}) to the corresponding Fermi-Dirac distribution function, $f_0$. On this 
regime, neutrinos are expected to \textit{relax} in a very short time scale and their distribution 
function differs from its equilibrium value only by a small anisotropic factor. In terms of a 
Legendre expansion of $f$, the diffusion approximation can be written as
\begin{align}\label{eq:DiffusiveApprox}
  f(E_\nu)\simeq f_0(E_\nu)+\mu f_1(E_\nu),
\end{align}
\noindent where $\mu$ is the cosine of the angle between the neutrino propagation direction and the 
radial vector, and $|f_1(E_\nu)|\ll 1$. 

All the quantities we are interested in in the study of neutrino transport will be defined in terms 
of $f$ and its angular moments of the form $\frac{1}{2}\int_{-1}^1d\mu \mu^i f$ 
\cite{Bruenn1985, Cooperstein1992, Pons1999}. With the use of the approximation given by Eq.
(\ref{eq:DiffusiveApprox}), it is possible to write the angular moments of Eq. (\ref{eq:ColInt_B}) 
in a very convenient form. The $0$-th moment will be associated with neutrino source terms 
\cite{Pons1999} and, since we are assuming the detailed balance of Eqs. (\ref{eq:DetBalance}), it is 
identically zero. The $1$st moment, on the other hand, contains the traditional \textit{opacities} 
and it is given by 
\begin{align}\label{eq:Q1}
	\frac{1}{2}\int_{-1}^{1}\mu B[f]d\mu&=-\frac{f_1(E_\nu)}{3}\left(j_a+\frac{1}{\lambda_a}
  +\kappa_1\right),
\end{align}
\noindent where 
\begin{align}\label{eq:kappa_0}
	\kappa_1=&\frac{1}{(2\pi)^2}\int_0^\infty dE_\nu^\prime E_\nu^{\prime 2}\left\{
    \frac{\left[1-f_0(E_\nu^\prime)\right]}{\left[1-f_0(E_\nu)\right]}R^{out}_0\right.\nonumber\\
		&\left.-\frac{f_0(E_\nu)}{f_1(E_\nu)}\frac{f_1(E_\nu^\prime)}{f_0(E_\nu^\prime)}
		R_1^{out}\right\}.
\end{align}
\noindent We have defined here the Legendre moments of the scattering kernel in terms of the 
scattering angle $\theta$
\begin{align}
  R_l^{out}=\int_{-1}^1d\cos\theta P_l(\cos\theta)R^{out}(E_\nu,E_\nu^\prime,\cos\theta),
\end{align}
\noindent with $\cos\theta=\mu\mu^\prime+\sqrt{(1-\mu^2)(1-\mu^{\prime 2})}\cos\phi$, where $\phi$ 
is the azimutal angle of one neutrino with respect to each other. 

The first term between the curly braces in Eq. (\ref{eq:kappa_0}) gives the inverse neutrino 
scattering mean free path when neutrinos are in thermodynamic equilibrium with the rest of the 
matter, and we denote it here by $1/\lambda_s$: 
\begin{align}\label{eq:lambda_scatt}
  \frac{1}{\lambda_s}=&\frac{\left[1-f_0(E_\nu)\right]^{-1}}{(2\pi)^2}\int_0^\infty dE_\nu^\prime 
	E_\nu^{\prime 2}\left[1-f_0(E_\nu^\prime)\right]\nonumber\\
	&\times\int_{-1}^1 d \cos\theta	R^{out}(E_\nu,E_\nu^\prime,\cos\theta).
\end{align}
\noindent Eq. (\ref{eq:lambda_scatt}) is just Eq. (\ref{eq:CollisionInt}) with $f_0$ for the 
neutrino distribution function and it can be written without one explicitly defining the scattering 
kernel $R^{out}$.

When the scattering process is isoenergetic (meaning that neutrinos and matter do not exchange 
energy), regardless the actual neutrino distribution function, Eq. (\ref{eq:kappa_0}) specializes to
\begin{align}\label{eq:IS}
  \kappa_1^{\mbox{\tiny{IS}}}=&\frac{E_\nu^2}{(2\pi)^2}\int_{-1}^{1}d \cos\theta\nonumber\\
	&\times (1-\cos\theta)R^{out}(E_\nu,E_\nu,\cos\theta),
\end{align}
\noindent which is a common form of scattering opacity used in different applications.
\noindent In a general situation, however, $\kappa_1$ depends explicitly on $f_1$, which is,
\textit{a priori} unknown. 

From the energy-dependent diffusion coefficient 
$D(E_\nu)=\left(j_a+1/\lambda_a+\kappa_1\right)^{-1}$, we define the energy-averaged coefficients 
\cite{Pons1999}
\begin{align}\label{eq:Dndef}
  D_n=&\int_0^\infty dx x^n f_0(E_\nu)[1-f_0(E_\nu)]D(E_\nu),
\end{align}
\noindent where $x=E_\nu/T$. In particular, the Rosseland energy-averaged neutrino mean free path 
\cite{Sawyer1979, Reddy1998} can be defined as
\begin{align}\label{eq:Lambda_Ross}
  \lambda_R=\frac{D_4}{\int_0^\infty dx x^4 f_0(E_\nu)[1-f_0(E_\nu)]}.
\end{align}

%%%%%%%%%%%%%%%%%%%%%%%%%%%%%%%%%%%%%%%%%%%%%%%%%%%%%%%%%%%%%%%%%%%%%%%%%%%%%%%%%%%%%%%%%%%%%%%%%%%% SEC
% NEUTRINO INTERACTIONS                                                                            % NEUTRINO INT.
%%%%%%%%%%%%%%%%%%%%%%%%%%%%%%%%%%%%%%%%%%%%%%%%%%%%%%%%%%%%%%%%%%%%%%%%%%%%%%%%%%%%%%%%%%%%%%%%%%%%

\section{Neutrino interactions}
All neutrino processes under consideration are listed on Table \ref{table:cconsts}. Electron 
neutrino absorption on quarks $d$ and $s$ involve the exchange of a $W$ boson, while neutrino 
scattering on quarks or on leptons may involve either the exchange of $W$ or $Z$ bosons. At the 
energy density regime we are interested in, the neutrino energy is always much smaller than the $W$ 
and $Z$ masses and the interactions may be described in terms of current-current couplings 
\cite{Tubbs1975,Iwamoto1982,Reddy1998}. Weak charged currents are associated to both neutrino 
absorption and neutrino scattering on leptons of the same generation, while weak neutral currents 
take part in neutrino scattering on either leptons or quarks. Given that it is possible to write the 
charged current contribution of a scattering process in terms of a neutral current (and vice-versa), 
we may write the current-current interaction Lagrangian in a combined form as
\begin{align}\label{eq:Lint}
	\mathcal{L}_{int}=&\frac{G_F}{\sqrt{2}}\left[\bar{u}(\nu)\gamma_\mu(1-\gamma_5)u(4)\right]
  \nonumber\\
	&\times\left[\bar{u}(2)\gamma^\mu(C_A-C_V\gamma_5)u(3)\right] + \mbox{H.C.},
\end{align}
\noindent with $C_A$ and $C_V$ being the appropriate axial and vector coupling constants to be read 
from Table \ref{table:cconsts} and $G_F$ is the Fermi weak coupling constant 
($G_F/c\hbar^3=1.664\times 10^{-5}$GeV$^{-2}$).

The transition rates $W_{if}$ appearing in Eq. (\ref{eq:CollisionInt}) are obtained from the matrix 
element by means of Femi's golden rule:
\begin{equation}
  W_{if}=(2\pi)^4\delta^4\left(p_\nu+p_2-p_3-p_4\right)\frac{\left\langle |\mathcal{M}|^2 
  \right\rangle}{2^4E_\nu E_2E_3E_4},
\end{equation}
\noindent where $\left\langle |\mathcal{M}|^2 \right\rangle$ denotes the squared matrix element 
summed over final spins and averaged over the initial spins. From Eq. (\ref{eq:Lint}) we may then 
derive an expression for $W_{if}$ also valid for both absorption and scattering processes:
\begin{align}\label{eq:Wfi}
  W_{fi}=&\frac{G_F^2}{E_\nu E_2E_3E_3}\left[(C_V+C_A)^2(p_\nu\cdot p_2)(p_3\cdot p_4)\right.
    \nonumber\\
		&+(C_V-C_A)^2(p_\nu\cdot p_3)(p_2\cdot p_4)\nonumber\\
		&\left.-(C_V^2-C_A^2)m_2 m_3 (p_\nu\cdot p_4)\right]\nonumber\\
		&\times(2\pi)^4\delta^4\left(p_\nu+p_2-p_3-p_4\right),
\end{align}
\noindent where $m_i$ is the mass of the corresponding particle.

The explicit form of the expressions used for the inverse mean free paths and for the scattering 
kernels can be found on appendices \ref{sec:ap_mfp} and \ref{sec:scatt_kernel}, respectively.

\begin{table}
	\caption{Vector and axial vector coupling constants for the charged and neutral currents under 
  consideration.	We use the values $\cos\theta_C=0.973$ and $\sin^2\theta_W=0.231$. The 
  corresponding scattering of antineutrinos
	involves only the substitution $C_{\mbox{\scriptsize{A}}}\rightarrow -C_{\mbox{\scriptsize{A}}}$.}
	\label{table:cconsts}
	\centering
		\begin{tabular*}{\linewidth}{l @{\extracolsep{\fill}} cc}
		\\
		\hline\hline
		$\nu+2\rightarrow 3+4$ & $C_{\mbox{\scriptsize{V}}}$ & $C_{\mbox{\scriptsize{A}}}$\\
		\hline\\
		$\nu_e+d\rightarrow u+e^-$   & $\cos\theta_C$ & $\cos\theta_C$ \\
		$\nu_e+s\rightarrow u+e^-$   & $\sin\theta_C$ & $\sin\theta_C$ \\
		\\
		$\bar{\nu}_e+u\rightarrow d+e^+$   & $\cos\theta_C$ & $-\cos\theta_C$ \\
		$\bar{\nu}_e+u\rightarrow s+e^+$   & $\sin\theta_C$ & $-\sin\theta_C$ \\
		\\
		$\nu_l+u\rightarrow u+\nu_l$ & $\phantom{- }\frac{1}{2}-\frac{4}{3}\sin^2\theta_W$ & $\phantom{- }\frac{1}{2}$\\
		$\nu_l+d\rightarrow d+\nu_l$ & $-\frac{1}{2}+\frac{2}{3}\sin^2\theta_W$ & $-\frac{1}{2}$\\
    $\nu_l+s\rightarrow s+\nu_l$ & $-\frac{1}{2}+\frac{2}{3}\sin^2\theta_W$ & $-\frac{1}{2}$\\
    $\nu_e+e^-\rightarrow e^-+\nu_e$&\phantom{- }$\frac{1}{2}+2\sin^2\theta_W$ & $\phantom{- }\frac{1}{2}$\\				
    $\nu_{\mu,\tau}+e^-\rightarrow e^-+\nu_{\mu,\tau}$   & $-\frac{1}{2}+2\sin^2\theta_W$ & $-\frac{1}{2}$\\
		\\
		\hline
		\end{tabular*}
\end{table}

%%%%%%%%%%%%%%%%%%%%%%%%%%%%%%%%%%%%%%%%%%%%%%%%%%%%%%%%%%%%%%%%%%%%%%%%%%%%%%%%%%%%%%%%%%%%%%%%%%%% SEC
% QUARK MATTER EOS                                                                                 % QUARK EOS
%%%%%%%%%%%%%%%%%%%%%%%%%%%%%%%%%%%%%%%%%%%%%%%%%%%%%%%%%%%%%%%%%%%%%%%%%%%%%%%%%%%%%%%%%%%%%%%%%%%%

\section{Quark matter equation of state}
We describe quark matter constituted of \textit{u}, \textit{d} and \textit{s} quarks, plus electrons 
and electron-neutrinos in terms of the MIT bag model \cite{Farhi1984}. The thermodynamics follows from 
the Grand potential (per unit volume)
\begin{align}
  \Omega = \sum_{i=u,d,s,e,\nu}\Omega_{i},
\end{align}
\noindent where $\Omega_{i}$ is the thermodynamic potential for a gas of ideal fermions
\begin{align}\label{eq:bag_omega}
  \Omega_{i}(T,\mu_i)=-\frac{g_i T}{2\pi^2}\int dk k^2 
  \ln\left[1+e^{-\left(E_k-\mu_i\right)/T}\right].
\end{align}
Antiparticles are included through $\Omega_{i}(T,-\mu_i)$ and all derived thermodynamic quantities 
must be understood as net quantities, containing the contributions of both particles and 
antiparticles. 

The inclusion of the vacuum energy density $B$ (the bag constant), requires that both the pressure 
and energy density are modified according to

\begin{align}
  P&=-\Omega-B\label{eq:pbag},\\
	\rho&=\Omega+\sum_i\left( Ts_i+\mu_i n_i\right) + B\label{eq:rhobag},
\end{align}
\noindent where the entropy density, $s_i$, and the particle number densities, $n_i$, are derived 
from $\Omega$ by the usual thermodynamic relations.

The five chemical potentials ($\mu_u,\mu_d,\mu_s,\mu_e,\mu_{\nu_e}$) at given temperature, baryon 
number density, $n_B=(n_u+n_d+n_s)/3$, and fixed net electron-type lepton fraction, 
$Y_L=(n_e+n_{\nu_e})/n_B$, are determined by means of the conditions of $\beta$-equilibrium
\begin{align}
  \mu_d&=\mu_u+\mu_e-\mu_{\nu_e},\\
	\mu_s&=\mu_d,
\end{align}
\noindent and electric charge neutrality
\begin{align}
  n_e&=\frac{2}{3}n_u-\frac{1}{3}\left(n_d+n_s\right).
\end{align}

\begin{figure}[t]
\centering
\includegraphics[scale=0.8]{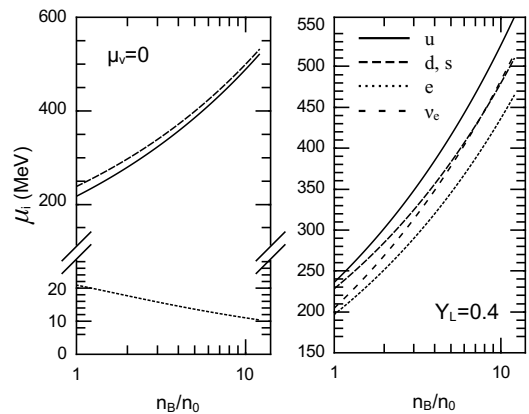}
\caption{\label{fig:1}Equilibrium chemical potentials of quark matter constituted only of the three 
lightest quarks  at $T=30$MeV as a function of the baryon number density. The panel at the left 
corresponds to neutrino-free quark matter. The panel at the right represents quark matter with 
trapped neutrinos, corresponding to a total electron-type lepton fraction of $Y_L=0.4$. For this 
figure, $B=60$MeV/fm$^3$ and $m_s=150$MeV.}
\end{figure}

\begin{figure}[t]
\centering
\includegraphics[scale=0.8]{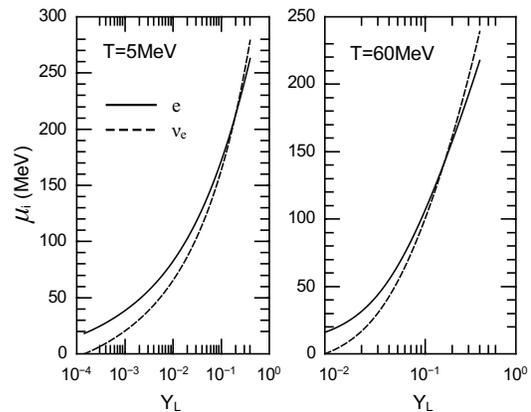}
\caption{\label{fig:2}Electron and electron-neutrino equilibrium chemical potentials in quark matter 
with trapped neutrinos at $n_B=2n_0$ for two different temperatures. The smallest value of $Y_L$ 
shown in this figure always represent neutrino-free quark matter, for which $Y_L=Y_e$. The largest 
value of $Y_L$ has been chosen to be $Y_L=0.4$ in both panels. Note how an increasing neutrino 
degeneracy is followed by an increasing electron degeneracy for fixed temperature and baryon density. 
For this figure, $B=60$MeV/fm$^3$ and $m_s=150$MeV. }
\end{figure}

When quark matter doesn't have trapped neutrinos, or when the trapped neutrinos are nondegenerate, 
the electron chemical potential is always much smaller than those of quarks. Quark matter in these 
circumstances requires only a small fraction of electrons to be electrically neutral. When the rate 
of occurrence of the weak interaction processes is high enough for neutrinos to become trapped and 
degenerate in quark matter, an increase in the abundance of electrons follows, as it can be seen 
from the behavior of the electron chemical potential in Fig. \ref{fig:1}. In fact, as shown in 
Fig. \ref{fig:1}, the degeneracy of electrons in quark matter follows closely that of neutrinos. 
This moderate uniformity between the electron and electron-neutrino chemical potentials, allows us 
to use the total electron-type lepton fraction $Y_L$ as an indicative of both the electron and 
electron-neutrino state of degeneracy, as we show in Fig. \ref{fig:2}.

QCD corrections to the thermodynamic potential of Eq. (\ref{eq:bag_omega}) for arbitrary
temperatures, quark masses and chemical potentials have been derived for the orders of $\alpha_c$ 
and $\alpha_c^{3/2}$ in perturbation theory \cite{Kapusta1979}. Closed-form expressions are known 
only on the approximate regimes. For degenerate massless quarks, one has, to first order in 
$\alpha_c=g^2/4\pi$ \cite{Kalashnikov1979} 
\begin{align}\label{eq:Omega_alphac}
  \Omega_{(2)}=\sum_{f=u,d,s}&\left[\frac{7}{60}\pi^2 T^4\left(\frac{50}{21}
  \frac{\alpha_c}{\pi}\right)\right.\nonumber\\
	&\left. + \left(\frac{1}{4\pi^2}\mu_f^4+\frac{1}{2}T^2\mu_f^2\right)\left(2
  \frac{\alpha_c}{\pi}\right)\right],
\end{align}
\noindent and the thermodynamic potential to second order is obtained through 
$\Omega \rightarrow \Omega+\Omega_{(2)}$. In the context of the MIT bag model, the strong coupling 
constant, $\alpha_c$, is regarded as a numeric constant. Together, $B$, $m_s$ and $\alpha_c$ 
constitute the three free parameters of the model.

Throughout this paper we use mainly the set of parameters $B=60$MeV/fm$^3$, $m_s=150$ MeV and $\alpha_c=0$, 
which falls inside the stability window of Ref. \cite{Farhi1984} and, consequently, describes absolutely 
stable (strange) quark matter, with an energy per baryon at zero pressure and temperature smaller than 
the neutron's mass.  However, we employ also other parameter sets, some of which represent standard 
(non-absolutely stable) quark matter. In particular, at the end of Sec. \ref{subsec:mfpana}, we show 
how each of $B$, $m_s$ and $\alpha_c$ 
affect the neutrino mean free paths through their influence on the equilibrium chemical potentials.

%%%%%%%%%%%%%%%%%%%%%%%%%%%%%%%%%%%%%%%%%%%%%%%%%%%%%%%%%%%%%%%%%%%%%%%%%%%%%%%%%%%%%%%%%%%%%%%%%%%% SEC
% RESULTS                                                                                          % RESULTS
%%%%%%%%%%%%%%%%%%%%%%%%%%%%%%%%%%%%%%%%%%%%%%%%%%%%%%%%%%%%%%%%%%%%%%%%%%%%%%%%%%%%%%%%%%%%%%%%%%%% 

\section{Results}

\begin{figure*}[tb]
\centering
\includegraphics[scale=0.8]{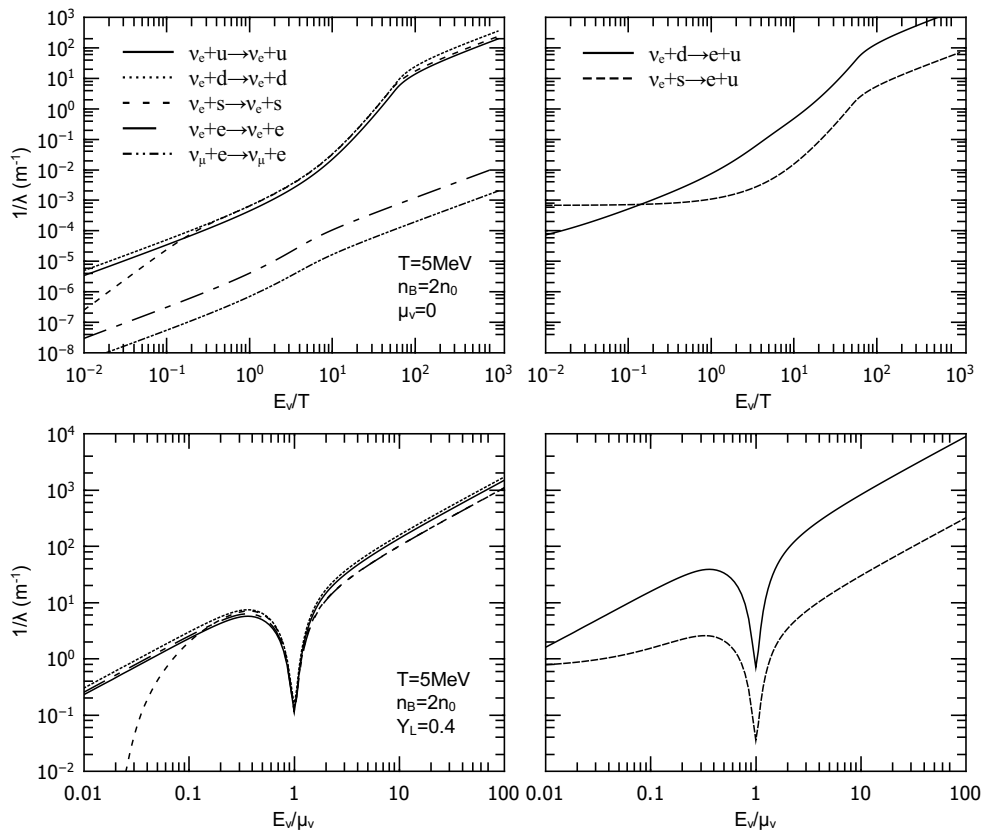}
\caption{\label{fig:3}Inverse neutrino mean free paths, Eq. (\ref{eq:ap_mpf_lambda}), in quark 
matter at two times nuclear saturation density and $T=5$MeV. At the upper panels, we have 
$\mu_{\nu_e}=0$ representing nondegenerate neutrinos. At the left, we have included 
muon neutrino-electron scattering for comparison. At the lower panels, $Y_L=0.4$ represents 
highly degenerate neutrinos and matter. Results for degenerate neutrinos have been divided by 
$[1-f_0(E_\nu)]$, so the curves can be better seen in the range $E_\nu<\mu_\nu$.}
\end{figure*}

\begin{figure*}[tb]
\centering
\includegraphics[scale=0.8]{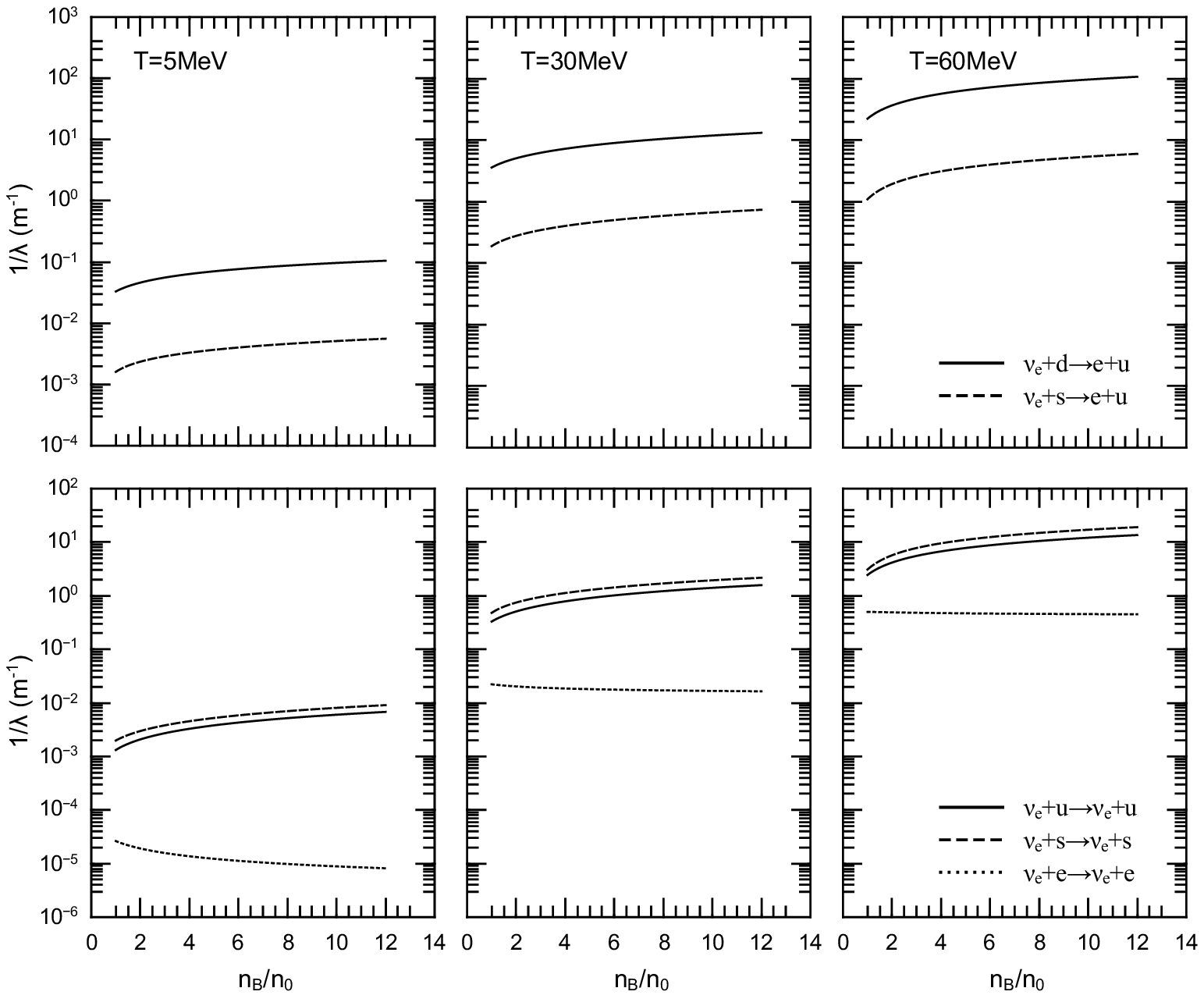}
\caption{\label{fig:4} Inverse mean free paths of nondegenerate neutrinos ($\mu_\nu=0$) in quark 
matter as a function of baryon density. On this figure, $E_\nu=3T$. In the lower panels, 
neutrino-down quark scattering was omitted since the corresponding curve superposes that of 
neutrino-strange quark scattering. }

\includegraphics[scale=0.8]{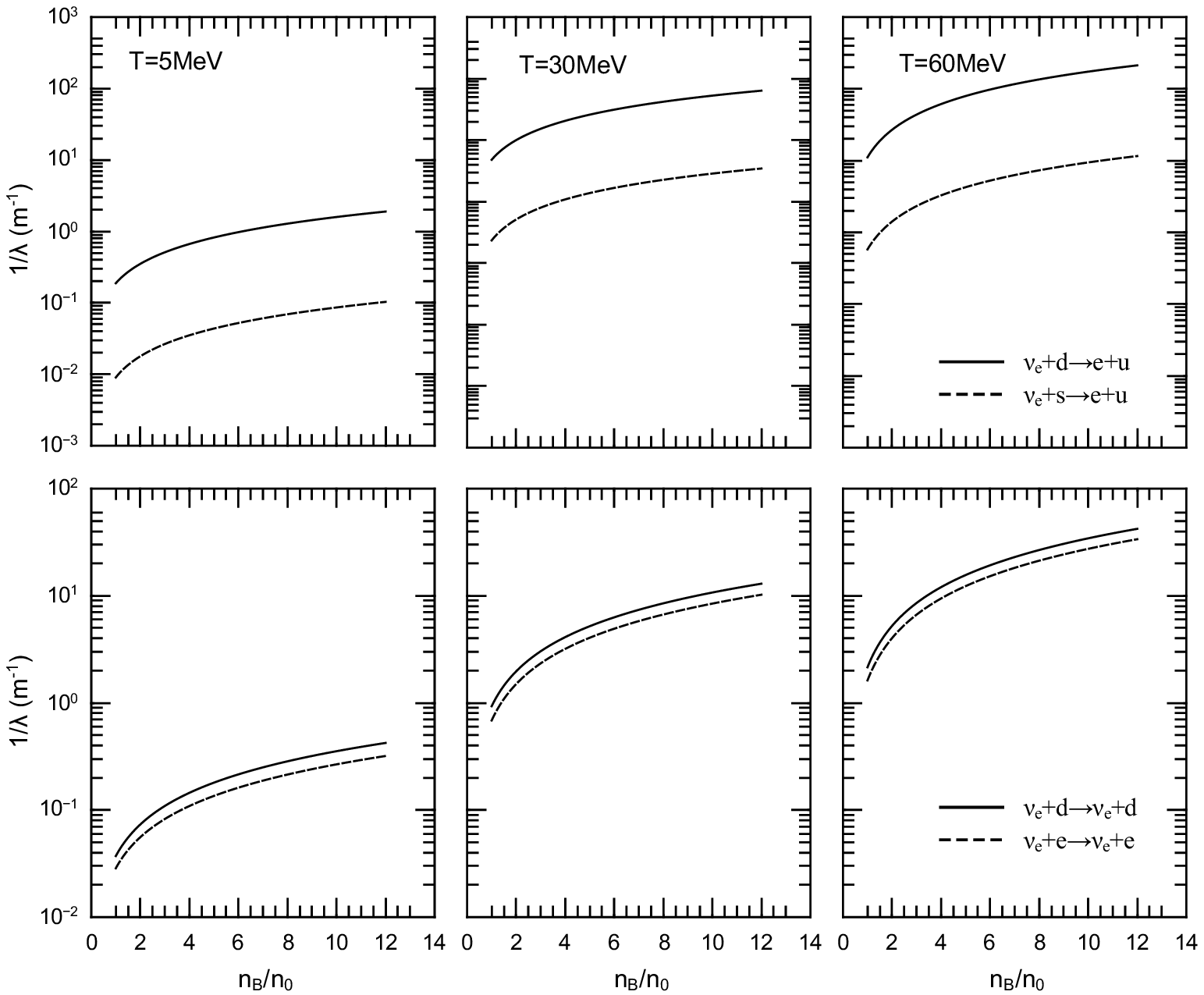}
\caption{\label{fig:5} Inverse mean free paths of degenerate neutrinos ($Y_L=0.4$) in quark matter 
as a function of baryon density. For this figure, $E_\nu=\mu_\nu$. The curves associated to neutrino 
scattering on $u$ and $s$ quarks lie between the curves shown in the lower panels and have been 
omitted. }
\end{figure*}

\begin{figure}
\centering
\includegraphics[scale=0.85,trim=0 0 10pt 0]{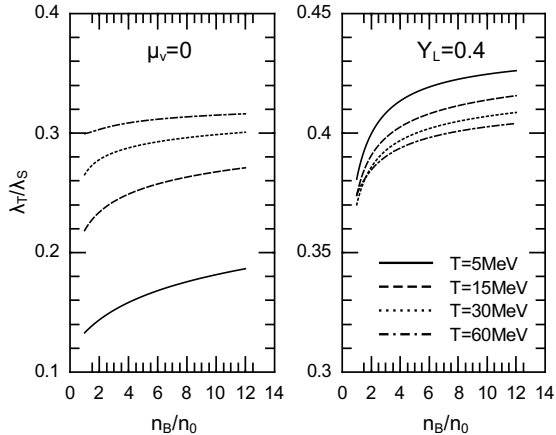}
\caption{\label{fig:6} Scattering contribution to the total Rosseland mean free path as a function 
of baryon density at several temperatures for both nondegenerate and degenerate neutrinos. All 
electron-neutrino processes of Table \ref{table:cconsts} have been included in the total opacity.}
\end{figure}

\begin{figure*}[htb]
\centering
\includegraphics[scale=0.8]{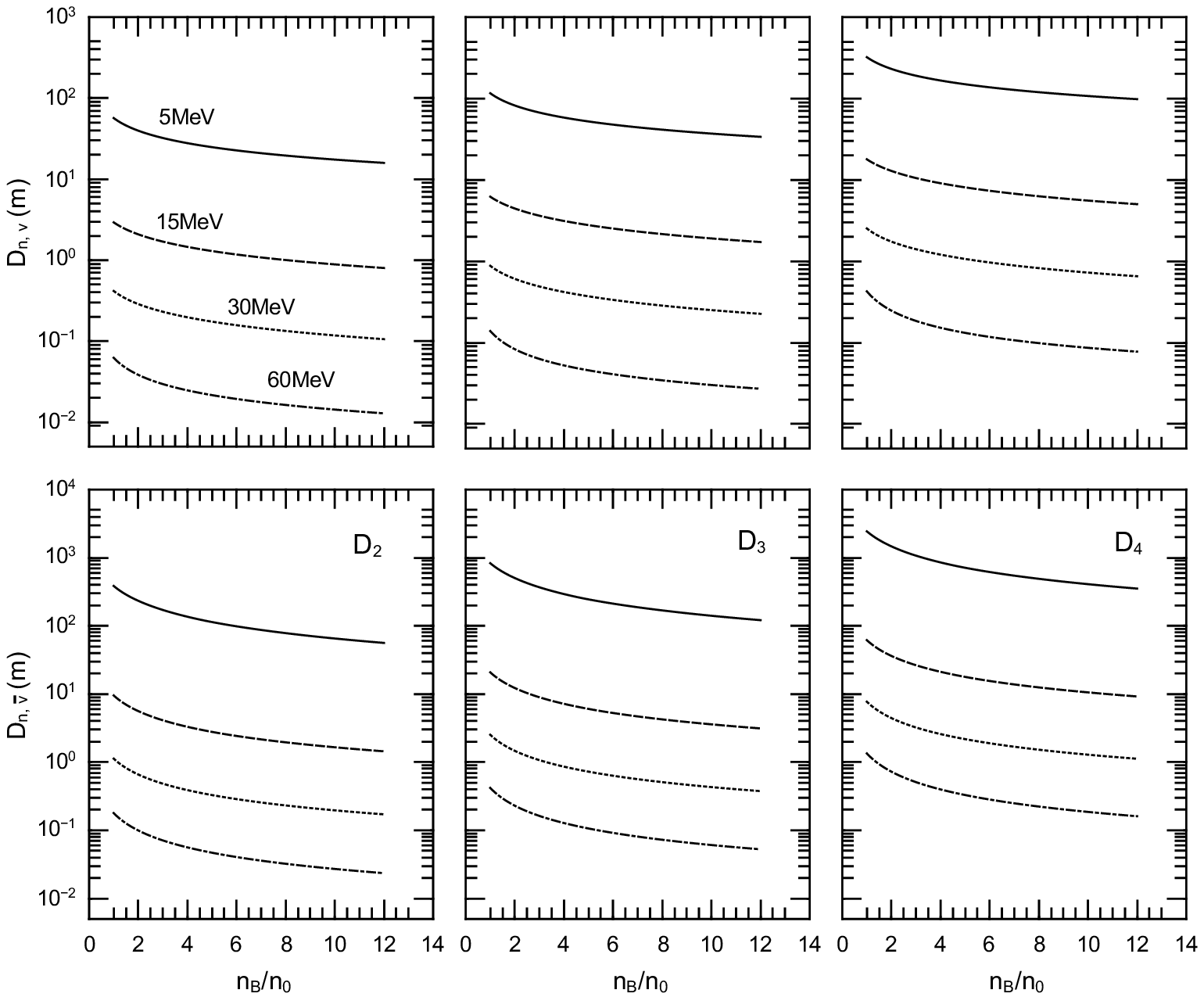}
\caption{\label{fig:7} Diffusion coefficients vs baryon density for nondegenerate ($\mu_\nu=0$) 
electron-neutrinos (upper panels)  and antineutrinos (lower panels) at different temperatures. The 
temperatures indicated in panel at the top left are the same for all the other panels.}
\includegraphics[scale=0.8]{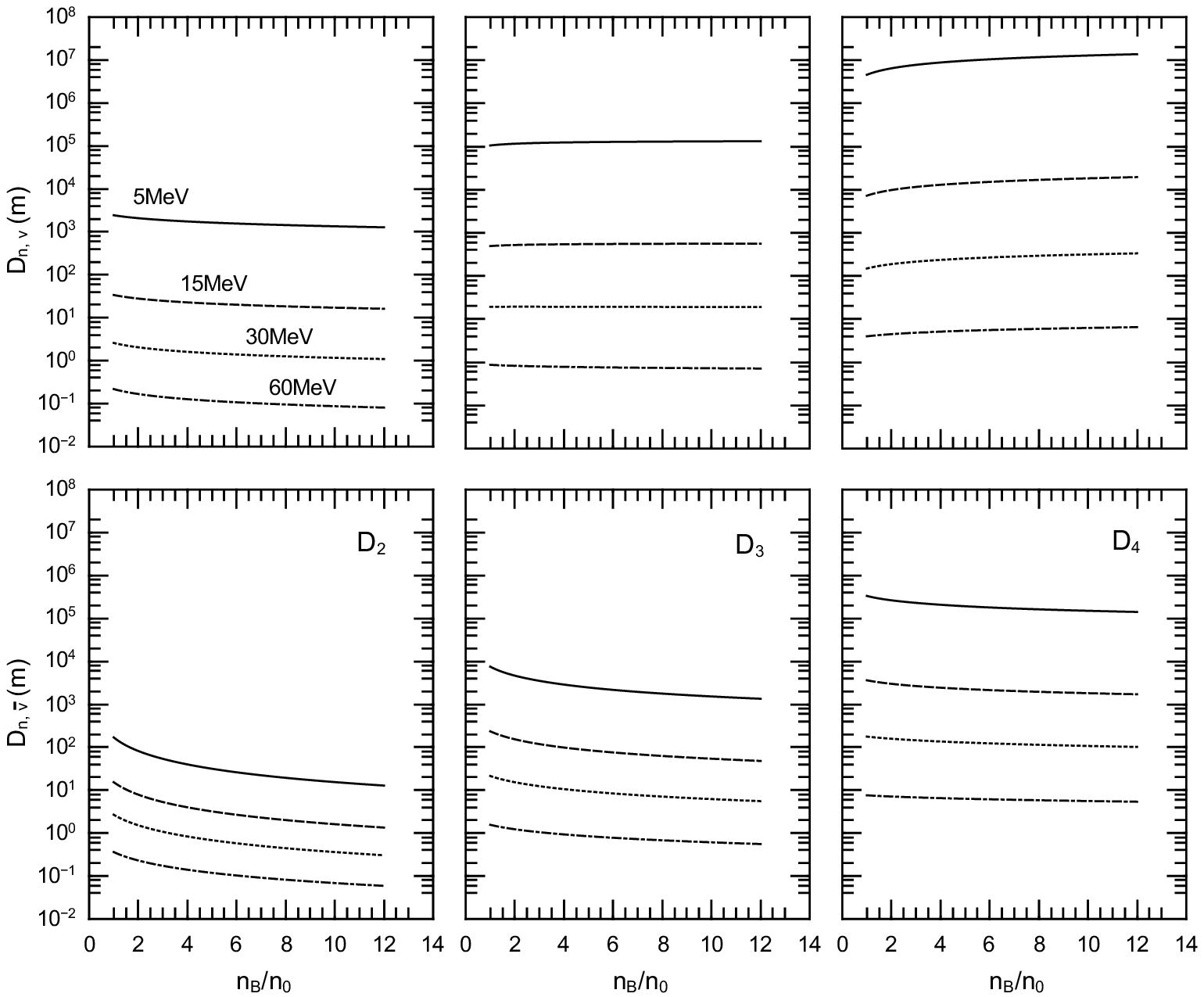}
\caption{\label{fig:8} Diffusion coefficients vs baryon density for degenerate ($Y_L=0.4$) 
electron-neutrinos (upper panels)  and antineutrinos (lower panels) at different temperatures. 
The temperatures indicated in panel at the top left are the same for all the other panels.}
\end{figure*}

\begin{figure}[htb]
\centering
\includegraphics[scale=0.85]{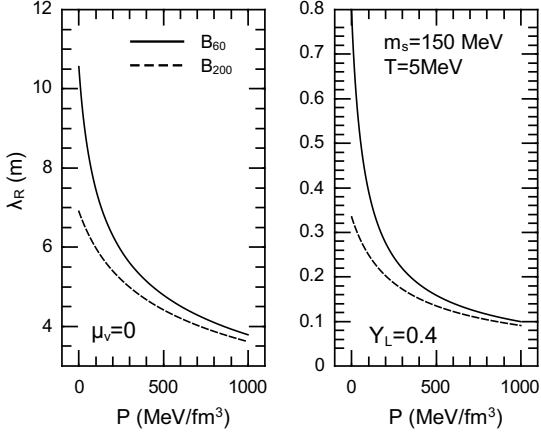}
\caption{\label{fig:9} Influence of the bag constant, $B$, on the
total Rosseland neutrino mean free path of degenerate neutrinos ($Y_L=0.4$) in quark matter. Here, 
$B_{60}$ and $B_{200}$ represent $B=60$MeV/fm$^3$ and $B=200$MeV/fm$^3$, respectively.}
\end{figure}

\begin{figure}[htb]
\centering
\includegraphics[scale=0.85]{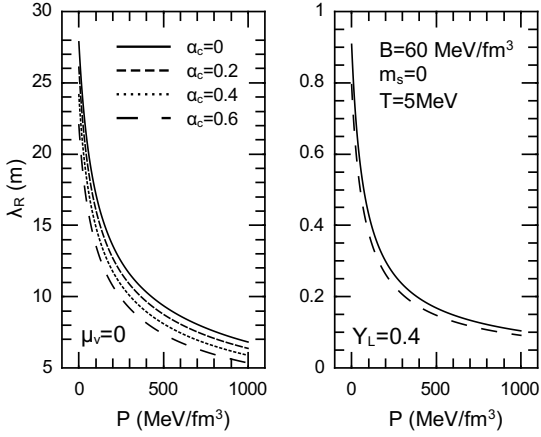}
\caption{\label{fig:10} Influence of the strong coupling constant, $\alpha_c$, on the
total Rosseland neutrino mean free path of nondegenerate (left panel) and of degenerate neutrinos 
(right panel, $Y_L=0.4$) in quark matter. In the panel at the right, only the cases $\alpha_c=0$ 
and $\alpha_c=0.6$ are considered. The curves corresponding to $\alpha_c=0.2$ and $\alpha_c=0.4$ 
lie between the two curves shown and have been omitted for clarity purposes.}
\end{figure}

\begin{figure}[htb]
\centering
\includegraphics[scale=0.85]{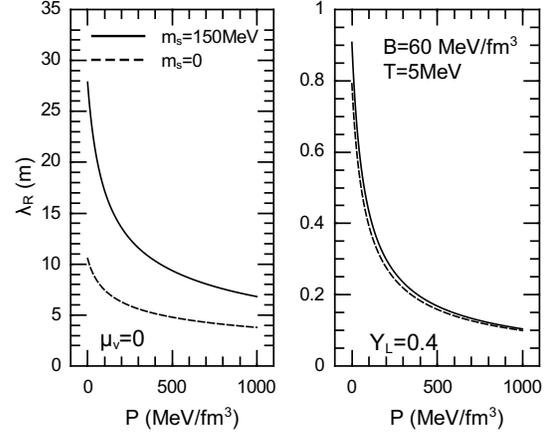}
\caption{\label{fig:11} Influence of the strange quark mass, $m_s$, on the
total Rosseland neutrino mean free path of nondegenerate neutrinos in quark matter with 
$B=60$MeV/fm$^3$.}
\end{figure}

\begin{figure}[htb]
\centering
\includegraphics[scale=0.85]{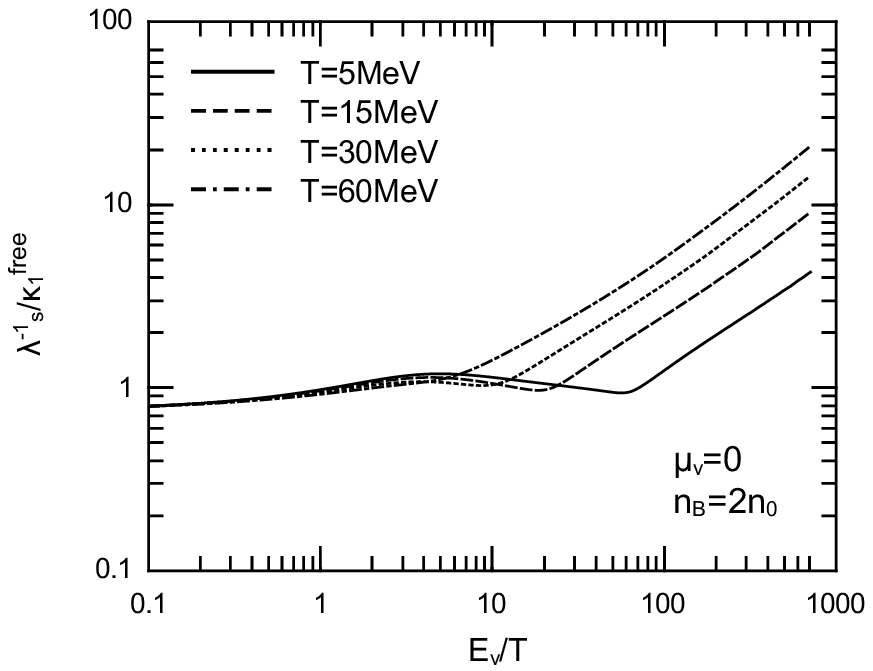}
\caption{\label{fig:12} Ratio between $1/\lambda_s$ given by Eq. (\ref{eq:lambda_scatt}) and 
$\kappa_1$, Eq. (\ref{eq:kappa_0}), with the limiting approximation $f_1(E_\nu)\propto f_0(E_\nu)$ 
for nondegenerate neutrinos ($\mu_{\nu_e}=0$).}
\includegraphics[scale=0.85]{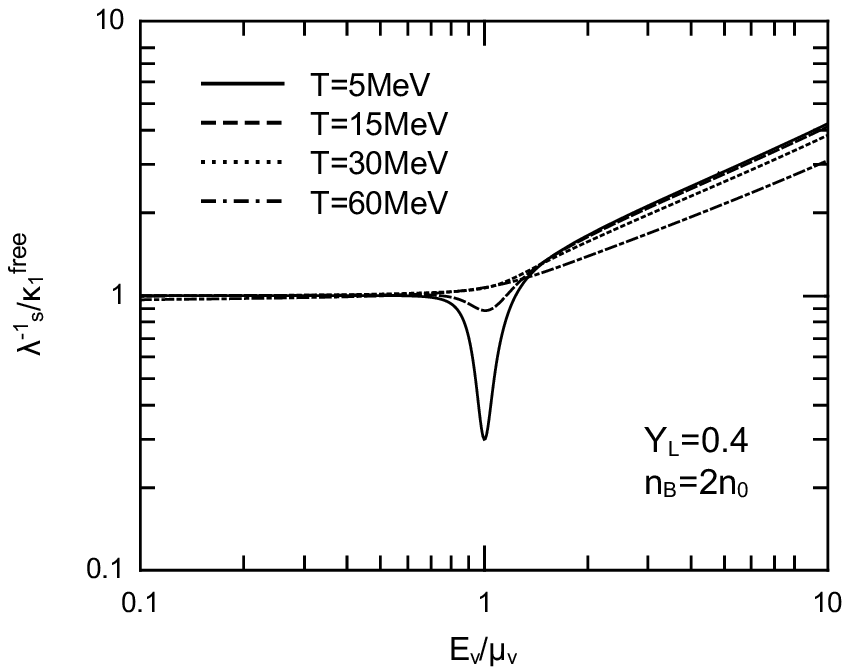}
\caption{\label{fig:13} The same as Fig. \ref{fig:12} but for degenerate neutrinos ($Y_L=0.4$).}
\end{figure}

\begin{figure*}[htb]
\centering
\includegraphics[scale=0.8]{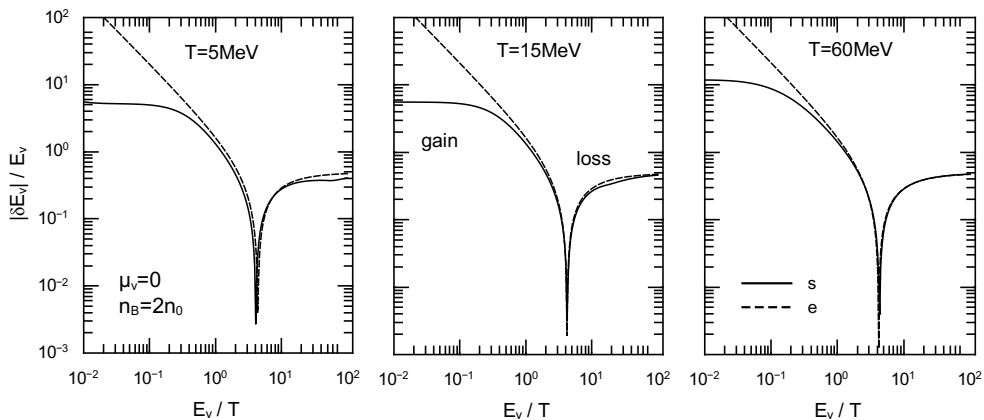}
\caption{\label{fig:14} Fractional mean energy transfer between nondegenerate neutrinos 
($\mu_\nu=0$) and matter at different temperatures. The energy transfer vanishes on average at 
$E_\nu\simeq 4T$.}
\end{figure*}
\begin{figure*}[htb]
\centering
\includegraphics[scale=0.8]{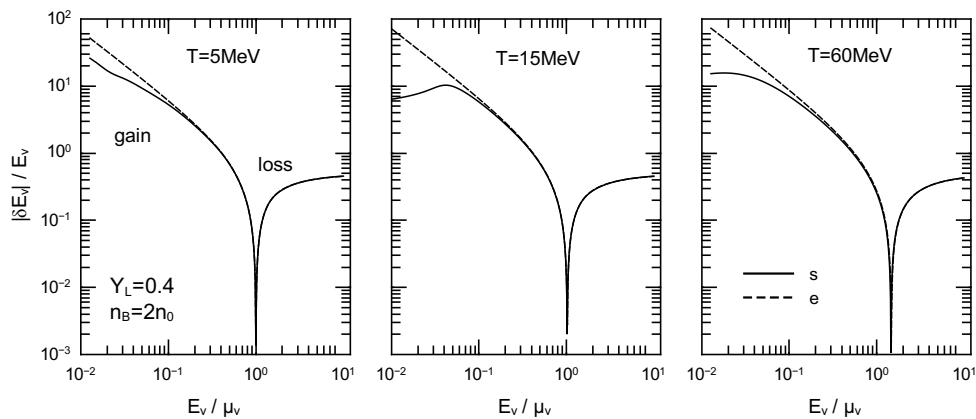}
\caption{\label{fig:15} Fractional mean energy transfer between degenerate neutrinos and quark 
matter at different temperatures. For very high degeneracy, the energy transfer vanishes on average 
at $E_\nu\simeq \mu_\nu$. For $T=60$MeV, both neutrinos and quark matter are slightly less 
degenerate and the energy transfer vanishes on average at $E_\nu$ between $\mu_\nu$ and $6T$. The 
curves associated with neutrino-up quark and neutrino-down quark scatterings are identical to the 
curves shown for neutrino-electron scattering and have thus been omitted. }
\end{figure*}

\begin{figure}[htb]
\centering
\includegraphics[scale=0.8]{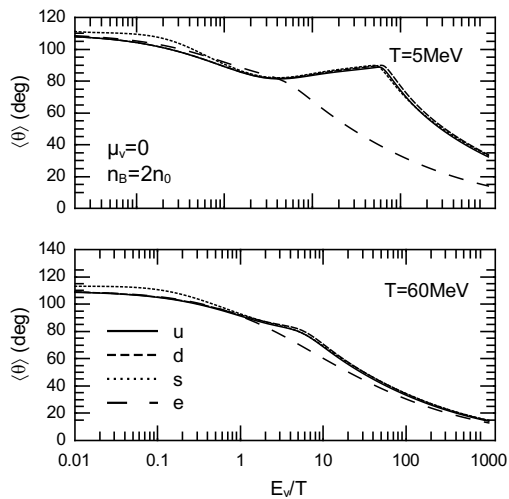}
\caption{\label{fig:16}Mean scattering angle $\langle \theta\rangle$ of nondegenerate neutrinos in 
quark matter. Here, $u$, $d$, $s$ and $e$ indicate the target particle.}
\end{figure}

\begin{figure}[htb]
\centering
\includegraphics[scale=0.8]{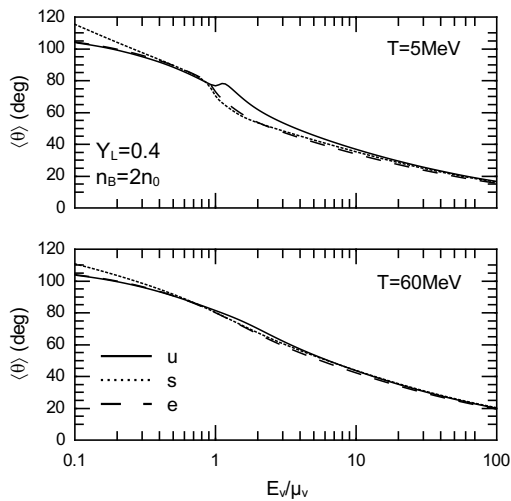}
\caption{\label{fig:17}The same as Fig. \ref{fig:16} but for $Y_L=0.4$.}
\end{figure}

\begin{figure}[htb]
\centering
\includegraphics[scale=0.8]{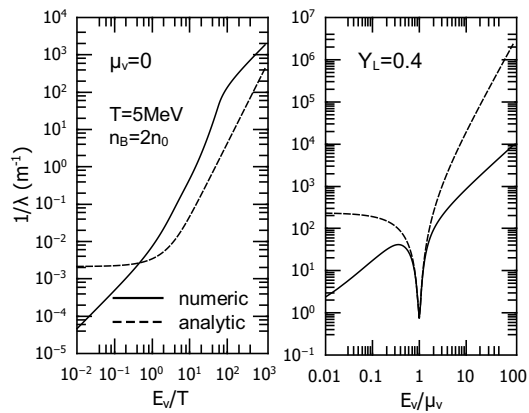}
\caption{\label{fig:18} Electron-neutrino inverse absorption mean free paths in quark matter at two 
times nuclear saturation density. Dashed lines indicate the approximations given in 
Eqs. (\ref{eq:Iw1982_abs_Dnu}) (right panel) and (\ref{eq:Iw1982_abs_NDnu}) (left panel). }
\end{figure}

\begin{figure*}[htb]
\centering
\includegraphics[scale=0.75]{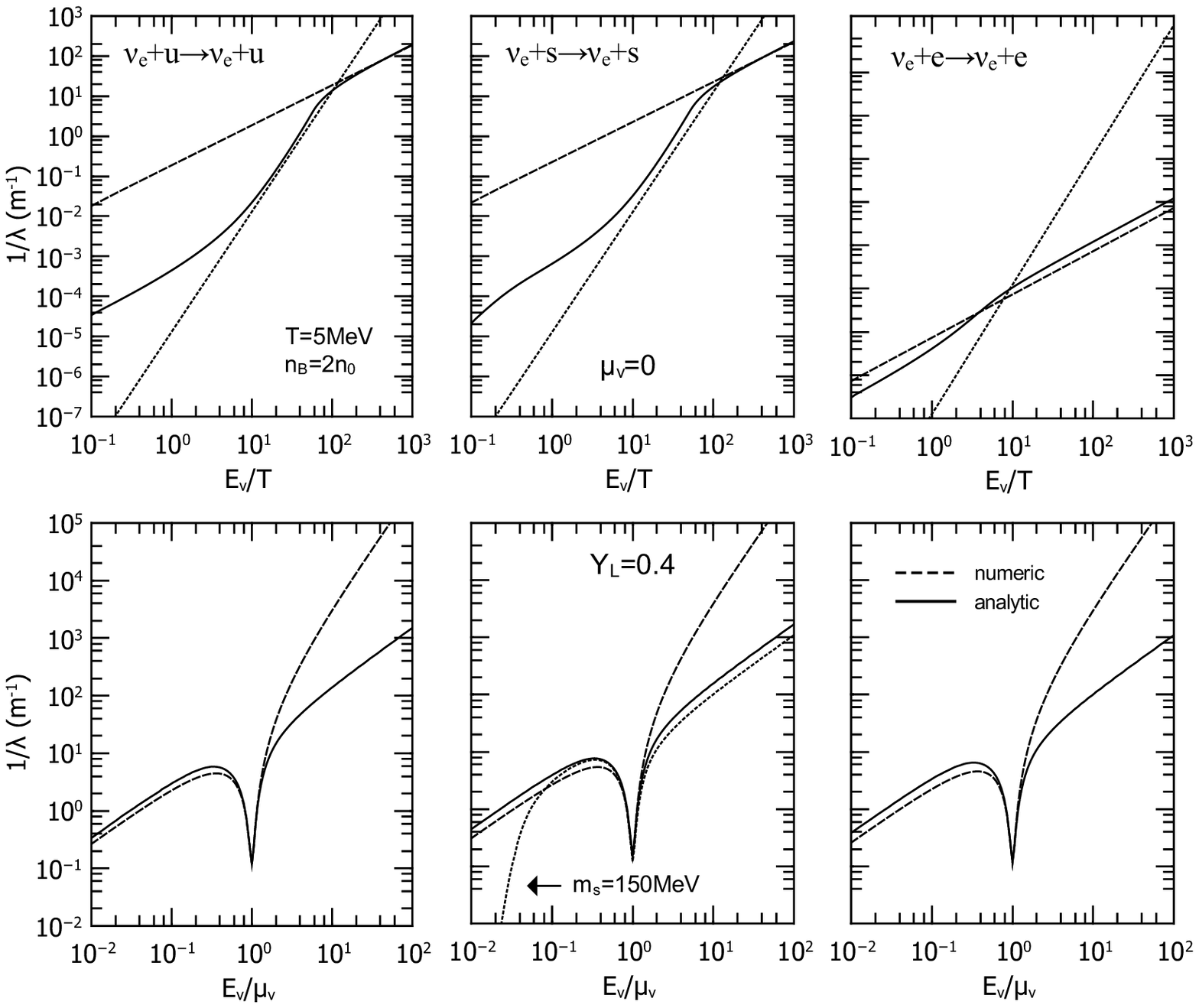}
\caption{\label{fig:19} Neutrino inverse scattering mean free paths. Upper 
panels: dashed lines indicate the approximations given in Eq. (\ref{eq:Iw1982_scatt_NDnu_high}) 
and dotted lines are the results of Eq. (\ref{eq:Iw1982_scatt_NDnu_low}). Lower panels: dashed lines 
indicate the results of Eq. (\ref{eq:Iw1982_scatt_Dnu}). For comparison purposes, all quarks masses 
have been neglected in the numeric integrations. In the lower panel on the middle, the dotted line 
corresponds to the numeric result when $m_s=150$MeV.}
\includegraphics[scale=0.75]{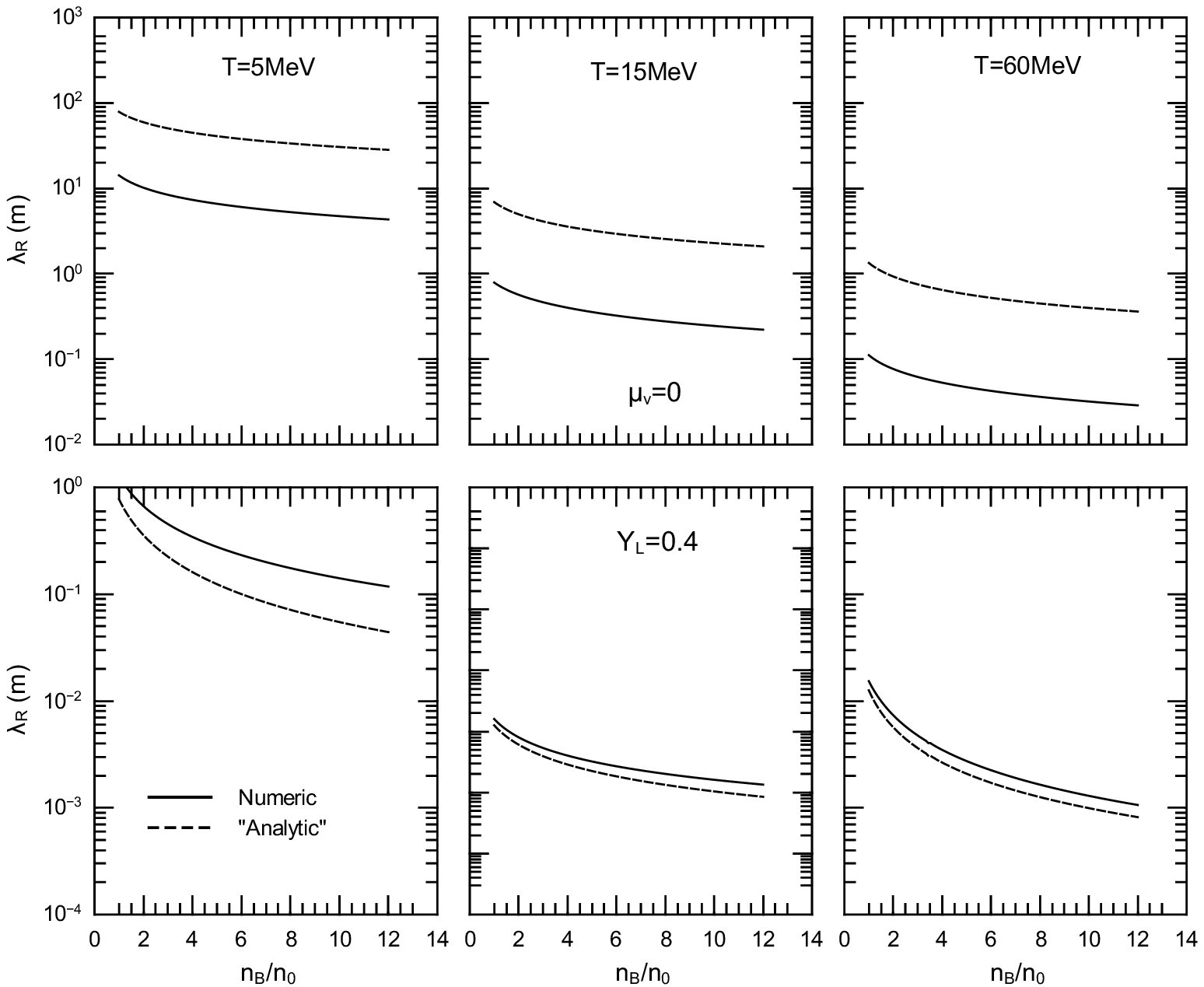}
\caption{\label{fig:20} Rosseland mean free paths for neutrinos in quark matter at different 
temperatures. ``Analytic" results mean numeric integration over the incident neutrino energies using 
the energy-dependent approximations of 
Eqs. (\ref{eq:Iw1982_abs_Dnu})-(\ref{eq:Iw1982_scatt_NDnu_low}). }
\end{figure*}

%%%%%%%%%%%%%%%%%%%%%%%%%%%%%%%%%%%%%%%%%%%%%%%%%%%%%%%%%%%%%%%%%%%%%%%%%%%%%%%%%%%%%%%%%%%%%%%%%%%% SUBSEC
% MFP Dn                                                                                           % MFP Dn
%%%%%%%%%%%%%%%%%%%%%%%%%%%%%%%%%%%%%%%%%%%%%%%%%%%%%%%%%%%%%%%%%%%%%%%%%%%%%%%%%%%%%%%%%%%%%%%%%%%%

\subsection{Neutrino inverse mean free paths and diffusion coefficients}\label{subsec:mfpana}

 Our results are shown separately in terms of either nondegenerate or highly degenerate
neutrinos. To nondegenerate neutrinos we associate a chemical potential $\mu_\nu=0$. Degenerate 
neutrinos, on the other hand, must have $\mu_\nu\gg T$. We represent the latter category in terms of 
a high net electron-neutrino lepton fraction $Y_L=(n_{e}+n_{\nu_e})/n_B$, where $n_{e}$ and 
$n_{\nu_e}$ represent, respectively, electron and electron-neutrino net number densities (number 
density of particles minus number density of the corresponding antiparticles) and $n_B$ is the 
baryon number density. 

In Fig. \ref{fig:3} we show the inverse neutrino mean free paths associated to the six 
electron-neutrino processes listed in Table \ref{table:cconsts}. It must be noted that the factor 
$[1-f_0(E_\nu)]^{-1}$ present in Eq. (\ref{eq:lambda_scatt}) has been kept so the curves can be 
better distinguished in the region $E_\nu<T$.

When neutrinos are nondegenerate, neutrino-quark scattering dominates the scattering opacity with 
respect to neutrino-electron scattering by orders of magnitude. This discrepancy can be explained by 
the tiny density of electrons present in quark matter. The presence of trapped degenerate neutrinos 
also increases the electron density and hence the neutrino-electron contribution to the scattering 
opacity also increases, as can be seen in the lower panel at the left. In both the upper and the 
lower panels at the left in Fig. \ref{fig:3}, it can be seen that neutrino-strange quark scattering 
differs from the other scatterings mainly at lower neutrino energies. In the panels at the right in 
Fig. \ref{fig:3} we show the neutrino inverse mean free paths associated with absorption. 
In general, the process of neutrino absorption on $d$ quarks dominates the opacity. The contribution 
of neutrino absorption on $s$ quarks, on the other hand, tends to be smaller than that of 
scatterings, with exception of the very low neutrino energy regime, in which this process dominates 
the opacity. These results are in agreement with the results found in Ref. \cite{Steiner2001}.

The behavior of the neutrino inverse mean free paths with respect
to baryon number density for fixed neutrino energies is shown in Figs. \ref{fig:4} and \ref{fig:5}.
 It can be seen that the contribution of neutrino-electron scattering is
always small in comparison with that of neutrino-quark scattering when neutrinos are nondegenerate. 
It is only when neutrinos are trapped and highly degenerate, implying in an also higher abundance of 
electrons, that neutrino-electron scattering becomes an important contribution to the total opacity, 
as it is shown in the lower panels of Fig. \ref{fig:5}. It is noticed that, for fixed $T$, the 
inverse mean free paths vary within one order of magnitude in the baryon number density range of 
interest. In contrast, the temperature dependence is stronger, e.g. there is a variation of more 
than two orders of magnitude when $T$ goes from $5$ MeV to a value above $30$ MeV.

The combined scattering processes amount to at least 10\% of the total Rosseland mean free path, as 
shown in Fig. \ref{fig:6}. In the case of leptonized quark matter, neutrino scattering represents 
$\sim 40\%$ of the total neutrino opacity. It is only when neutrinos are less degenerate in colder 
quark matter that neutrino absorption becomes dominant. Even in this case, as we already mentioned, 
neutrino scattering will contribute to at least $\sim 10\%$ of the total opacity in lower density 
regions of the star where neutrinos are completely nondegenerate.

In Figs. \ref{fig:7} and \ref{fig:8} we show the diffusion coefficients
$D_2$, $D_3$ and $D_4$ for both electron-neutrinos and electron-antineutrinos. When neutrinos are 
nondegenerate, all diffusion coefficients decrease with increasing density. As it can be seen in the 
panels at the middle, the combined coefficient $D_3=D_{3,\nu_e}-D_{3,\bar{\nu}_e}$  will be 
negative. In Fig. \ref{fig:8} it can be seen that the behavior of $D_2$, $D_3$ and $D_4$ for 
degenerate neutrinos is not the same with respect to increasing density. Notoriously, $D_2$ 
decreases, while $D_4$ increases. This behavior can be anticipated by means of analytic 
approximations for degenerate neutrinos, as we show in Appendix \ref{Ap:Dn}. 

In Figs. \ref{fig:9} through \ref{fig:11} we analyze separately how $m_s$, $B$ and $\alpha_c$ 
indirectly influence the neutrino Rosseland mean opacity through their effects on the equilibrium 
composition of quark matter as a function of pressure for a given temperature. In 
Eq. (\ref{eq:pbag}), it is seen that, for a given $P$, a larger $B$ will imply in a larger absolute 
value of $\Omega$. For fixed $m_s$ and $\alpha_c$, it follows that the equilibrium chemical 
potentials will increase accordingly. As verified in Fig. \ref{fig:9}, this implies in smaller mean 
free paths. From Eq. (\ref{eq:Omega_alphac}), it follows that the same reasoning applies for 
increases in $\alpha_c$ and this is shown in Fig. \ref{fig:10}.
For last, we analyze the influence of the strange quark mass on the neutrino mean free paths. Since 
$m_s$ enters not only in the equation of state but also directly in the computation of the mean free 
paths in a nontrivial way, its influence is inferred directly from the numeric results shown in 
Fig. \ref{fig:11}, where it can be seen that an increase in $m_s$ leads to larger mean free paths.

It follows then that increases in both $B$ and $\alpha_c$ lead to decreases in the neutrino mean 
free paths. The effect of increasing $\alpha_c$, however, is distinct from that of increasing $B$ 
in the sense that its effect is more prominent at higher pressures. It can also be noted that the 
mean free paths of highly degenerate neutrinos in highly degenerate quark matter is less sensitive 
to changes in $m_s$ or $\alpha_c$, since in these situations the chemical potentials are already 
very high. 

%%%%%%%%%%%%%%%%%%%%%%%%%%%%%%%%%%%%%%%%%%%%%%%%%%%%%%%%%%%%%%%%%%%%%%%%%%%%%%%%%%%%%%%%%%%%%%%%%%%% SUBSEC
% ERROR ESTIMATE                                                                                   % ERROR
%%%%%%%%%%%%%%%%%%%%%%%%%%%%%%%%%%%%%%%%%%%%%%%%%%%%%%%%%%%%%%%%%%%%%%%%%%%%%%%%%%%%%%%%%%%%%%%%%%%%

\subsection{Error estimate associated with the disregard of the anisotropic contribution to the 
neutrino distribution function in scatterings }\label{subsec:kappa_error}

We have seen that scattering may represent a considerable fraction of the total neutrino opacity in 
quark matter. It is evident then that neutrino scattering must be treated with proper care. Now we 
wish derive an estimate of the error committed when the term containing $f_1$ in 
Eq. (\ref{eq:kappa_0}) is neglected and Eq. (\ref{eq:lambda_scatt}) is used as the neutrino inverse 
scattering mean free path. Even though we do not know an explicit form for $f_1$, it is expected to 
be much smaller than $f_0$ in magnitude on the regions of higher density, where we expect neutrinos 
to be almost in thermodynamic equilibrium with the remaining of the matter. In this case, the ratio 
$\lambda_s^{-1}/\kappa_1$ tends to unity, given that the term containing $f_1$ in 
Eq. (\ref{eq:kappa_0}) becomes vanishingly small compared to the first term and 
Eq. (\ref{eq:kappa_0}) tends to Eq. (\ref{eq:lambda_scatt}). On the other hand, on the regions of 
extremely small neutrino opacity, neutrinos will stream almost freely and, in the limit of free 
stream, $f_1(E_\nu)\sim 3f_0(E_\nu)$ and the ratio $\lambda_s^{-1}/\kappa_1$ deviates maximally
from the unity.  

A fairly general expression relating $f_1$ and $f_0$ on the surface of a stellar core may be written 
in the form \cite{Bruenn1985}
\begin{align}\label{eq:f1_surface}
  f_1(E_\nu)&=Kf_0(E_\nu),
\end{align}
\noindent where $K$ is responsible for smoothly shifting the neutrino flux from isotropic to 
radially outward as the optical depth increases. In fact, as the optical depth becomes larger, 
Eq. (\ref{eq:f1_surface}) will hold not only on the stellar surface but in a whole thick layer where 
neutrinos stream freely. In principle, $K$ should depend on the neutrino energy, however, to avoid 
further complications we consider it as a constant geometric factor. In this case, 
Eq. (\ref{eq:kappa_0}) reduces to
\begin{align}\label{eq:kmaxerr}
  \kappa^{\mbox{\scriptsize{free}}}_1&=\frac{1}{\lambda_s}-\frac{1}{(2\pi)^2}\int_0^\infty 
  dE_\nu^\prime E_\nu^2\int_{-1}^{1}d\cos\theta	\cos\theta R^{out}\nonumber\\
	&=\frac{1}{\lambda_s}-\frac{1}{(2\pi)^2}\int_0^\infty dE_\nu^\prime E_\nu^2 
  R^{out}_1(E_\nu,E_\nu^\prime),
\end{align}
\noindent and the ratio $\lambda^{-1}_s/\kappa^{\mbox{\scriptsize{free}}}_1$ represents a bound to the maximum 
deviation from unity possible. In other words, the ratio $\lambda^{-1}_s/\kappa_1$ will always be 
closer to $1$ than it is $\lambda^{-1}_s/\kappa^{\mbox{\scriptsize{free}}}_1$. This information 
can be used as an estimate of the largest possible error committed on the computation of the 
scattering contribution to the opacity when $\lambda^{-1}_s$ is used in the place of $\kappa_1$.

When neutrinos are nondegenerate, it can be seen in Fig. \ref{fig:12} that which between 
$\lambda^{-1}_s$ and $\kappa^{\mbox{\scriptsize{free}}}_1$ is greater depends on the incident 
neutrino energy. For $E_\nu < T$, $\lambda^{-1}_s$ tends to be $\sim 20\%$ smaller than 
$\kappa^{\mbox{\scriptsize{free}}}_1$. For $E_\nu \gg T$, we see that the contribution of the term 
containing $R^{out}_1$ in Eq. (\ref{eq:kmaxerr}) becomes larger and 
$\kappa^{\mbox{\scriptsize{free}}}_1$ becomes many times smaller than $\lambda^{-1}_s$ as the 
energy of the incident neutrino increases. In Fig. \ref{fig:13} we see that when neutrinos are 
degenerate, there's almost no distinction between the two calculated scattering opacities when 
$E_\nu\ll \mu_\nu$. However, for neutrinos with $E_\nu\simeq\mu_\nu$, there is a noticeable 
difference, that increases with increasing neutrino degeneracy. For instance, when $T=5$MeV, 
$\lambda^{-1}_s$ is $\sim 30\%$ smaller than $\kappa^{\mbox{\scriptsize{free}}}_1$ for $Y_L=0.4$. 
For higher neutrino energies with respect to $\mu_\nu$, $\lambda^{-1}_s$ becomes several times 
greater than $\kappa^{\mbox{\scriptsize{free}}}_1$, similarly to what happens when neutrinos are 
nondegenerate.

In diffusive transport schemes, one is generally interested in the energy-averaged diffusion 
coefficients as in Eqs. (\ref{eq:Dndef}) and (\ref{eq:Lambda_Ross}), whose integrands are weighted 
with the factor $f_0(1-f_0)$, which quickly cuts off the higher energies, and features a peak 
centered at $E_\nu=\mu_\nu$ when neutrinos are degenerate. It means that in practice, 
$\lambda^{-1}_s$ will give scattering opacities at most $\sim 20\%$ smaller than the the ones that 
would be obtained if ${\kappa}_1$ was used. We have seen that the scattering contribution to the 
total Rosseland mean opacity is in the range $10-30\%$ for temperatures in the range $5-60$MeV when 
neutrinos are nondegenerate (Fig. \ref{fig:6}). It follows then that the error committed in the 
total neutrino opacity, may be estimated to have an upper bound in the range $2-6\%$. In regions 
where neutrinos are very close to thermodynamic equilibrium, and the anisotropy in the neutrino 
distribution function is vanishingly small, this error is expected to be much smaller. 

The situation is very similar when neutrinos are degenerate. In this case, the largest errors will 
be associated to those shown in Fig. \ref{fig:13} for $E_\nu \simeq\mu_\nu$, which we see to be of 
about $30\%$ when $T=5$MeV and smaller for larger temperatures. In this case, neutrino scattering 
amounts to about $40\%$ of the total neutrino opacity (Fig. \ref{fig:6}) and the error committed 
may be estimated to have an upper bound in the range $12-4\%$ for temperatures increasing in the 
range $T=5-60$MeV. If neutrinos are not expected to be too strongly degenerate in quark matter with 
temperatures as low as $5$ MeV, then in this case the estimated upper bound for the error will be 
closer to $4\%$.

%%%%%%%%%%%%%%%%%%%%%%%%%%%%%%%%%%%%%%%%%%%%%%%%%%%%%%%%%%%%%%%%%%%%%%%%%%%%%%%%%%%%%%%%%%%%%%%%%%%% SUBSEC
% ENERGY TRANSFER / SCATT. ANGLE                                                                   % ET SCATT. ANGLE
%%%%%%%%%%%%%%%%%%%%%%%%%%%%%%%%%%%%%%%%%%%%%%%%%%%%%%%%%%%%%%%%%%%%%%%%%%%%%%%%%%%%%%%%%%%%%%%%%%%%

\subsection{Mean energy transfer and mean scattering angle}
Neutrino-electron scattering has long been known to be an important thermalizing agent in neutron 
star matter matter \cite{Tubbs1975,Tubbs1979}, while neutrino-baryon scattering is usually regarded 
as an isoenergetic process \cite{Bruenn1985,Pons1999}. In Ref. \cite{Reddy1997}, Prakash \& Lattimer 
work in detail neutrino scattering in neutron star matter with and without hyperons, nevertheless, 
no attention has been paid to the aspects of neutrino/matter energy exchange or neutrino mean 
scattering angles.

A detailed analysis of neutrino scattering in quark matter, however, is missing in the literature. 
Given that quarks are very degenerate in quark matter and considering, for instance, the large mass 
of the strange quark with respect to that of electrons, it remains the question of whether 
neutrino-quark scattering is similar to neutrino-baryon or to neutrino-electron scattering with 
respect to neutrino-matter energy transfer and mean scattering angles. In this section we show that 
neutrino-quark scattering is always similar to neutrino-electron scattering with respect to energy 
exchange and hence non-isoenergetic. Both the mean energy transfer and the mean scattering angle 
are defined in Appendix \ref{Ap:etransfer}.

The first thing to be noticed in Figs. \ref{fig:14} and \ref{fig:15}, is that neutrino-matter
energy exchange vanishes on average for a particular incident neutrino energy, $E_0$, depending on 
both the neutrino and matter state of degeneracy. For $E_\nu<E_0$, neutrinos tend to gain energy on 
average, while for $E_\nu>E_0$, they tend to lose energy. This is not to say that neutrinos with 
$E_\nu=E_0$ don't exchange energy with matter, but rather, that the net energy exchange vanishes on 
average. We see that the energy-exchange in neutrino-strange quark scattering is very similar to 
that in neutrino-electron scattering. The curves representing neutrino-up quark and neutrino-down 
quark scattering have been omitted since the difference between those and that of 
neutrino-electron scattering would be barely visible.

When neutrinos are nondegenerate, $E_0$ lies between $4T$ and $5T$. When neutrinos are degenerate, 
$E_0$ lies between $\mu_\nu$ and $6T$. The exact value depends on the state of degeneracy of the 
matter, which determines both the mean energy of the particles participating on the reaction and the 
phase space available for the scattered particles \cite{Tubbs1975,Tubbs1978,Tubbs1979}.

Regardless of the neutrino/matter particular state of degeneracy, it is a general result that highly 
energetic neutrinos ($E_\nu\gg E_0$) will lose on average half their initial energy in each 
scattering process, a result previously known for the case of neutrino-electron scattering when 
neutrinos are nondegenerate \cite{Bahcall1964,Tubbs1975}, and which we extend here for degenerate 
neutrinos as well. On the other hand, very low-energetic neutrinos can gain on average several times 
their initial energy in a given scattering process and it is in this energy regime that 
neutrino-strange quark and neutrino-electron scatterings differ. It can be seen that low-energetic 
neutrinos tend to gain more energy from electrons (and from up and down quarks) than from
strange quarks.

Figs. \ref{fig:16} and \ref{fig:17} show the neutrino mean scattering angle. Less energetic 
neutrinos will scatter on average by an angle of about $120$ degrees. In association with 
Figs. \ref{fig:14} and \ref{fig:15}, this indicates not only high energy transfer but also high 
momentum transfer. The mean scattering angle decreases as the energy of the incident neutrino
increases, as shown in Fig. \ref{fig:16}, meaning a more isotropic scenario. When $E_\nu$ approaches 
the target's chemical potential, we see a local peak in $\langle \theta\rangle$. More highly 
energetic neutrinos will on average scatter almost equally on any direction, resulting in small mean 
scattering angles.

%%%%%%%%%%%%%%%%%%%%%%%%%%%%%%%%%%%%%%%%%%%%%%%%%%%%%%%%%%%%%%%%%%%%%%%%%%%%%%%%%%%%%%%%%%%%%%%%%%%% SUBSEC
% ANAL. COMP.                                                                                      % AN. COMP.
%%%%%%%%%%%%%%%%%%%%%%%%%%%%%%%%%%%%%%%%%%%%%%%%%%%%%%%%%%%%%%%%%%%%%%%%%%%%%%%%%%%%%%%%%%%%%%%%%%%%

\subsection{Comparison with analytic approximations}

When the matter is known to be completely degenerate, it is possible to find approximate expressions 
for the neutrino mean free paths with respect to the processes of absorption and of scattering, 
considering that the momenta of all matter constituents are fixed at their Fermi surface values. 
Neutrinos, on their turn, are considered to be either completely degenerate or nondegenerate. 

For the absorption of degenerate neutrinos, it is found \cite{Sawyer1979, Iwamoto1982} 
\begin{align}\label{eq:Iw1982_abs_Dnu}
  	\frac{1}{\lambda_a^{\mbox{\scriptsize{D}}}}=&\frac{4G_F^2}{\pi^3}\cos^2\theta_c
    \frac{p_{F_u}^2p_{F_e}^3}{\mu_\nu^2}\left[(E_\nu-\mu_\nu)^2+(\pi T)^2\right]&\nonumber\\ 
		&\times\left[1+\frac{1}{2}\frac{p_{F_e}}{p_{F_u}}+\frac{1}{10}
    \left(\frac{p_{F_e}}{p_{F_u}}\right)^2\right],
\end{align}
\noindent where $p_{F_i}$ is the Fermi momentum of particle species $i$. 
Eq. (\ref{eq:Iw1982_abs_Dnu}) is suitable for the case $|p_{F_u}-p_{F_e}|\geq |p_{F_d}-\mu_\nu|$. 
In the opposite situation, i.e., when $|p_{F_u}-p_{F_e}|\leq |p_{F_d}-\mu_\nu|$, the appropriate 
mean free path is obtained through the replacements $p_{F_u}\rightarrow p_{F_d}$ and 
$p_{F_e}\rightarrow \mu_\nu$. For the absorption of nondegenerate neutrinos, it is found
\begin{align}\label{eq:Iw1982_abs_NDnu}
\frac{1}{\lambda_a^{\mbox{\scriptsize{ND}}}}=&\frac{16}{\pi^4}\alpha_c G_F^2\cos^2\theta_c 
  p_{F_u}p_{F_d}p_{F_e}\frac{E_\nu^2+\left(\pi T\right)^2}{1+e^{-E_\nu/T}}
\end{align}
\noindent The strong coupling constant, $\alpha_c$, 
makes it appearance in Eq. (\ref{eq:Iw1982_abs_NDnu}) through the lowest-order corrections to the 
quarks' chemical potentials due to quark-gluon strong interaction
\begin{align}\label{Eq:pF_alphac}
  \mu_i=\left(1+\frac{8}{3\pi}\alpha_c\right)p_{F_i},
\end{align}
\noindent necessary to make the reaction kinematically allowed, since the masses of the $u$ and $d$ 
quarks are small and the neutrino momentum is not expected to play any significant role in the 
momentum conservation when they are nondegenerate, and hence, excluded from the 
momentum-conservation delta function \cite{Iwamoto1982}.

The influence of the neutrino state of degeneracy on neutrino-electron scattering has been analyzed 
to great extent. The inverse neutrino mean free path for degenerate neutrinos for this process has 
been found to be approximately
\cite{Lamb1976,Goodwin1982,Iwamoto1982}
\begin{align}\label{eq:Iw1982_scatt_Dnu}
	\frac{1}{\lambda_s^{\mbox{\scriptsize{D}}}}=&\frac{n_i\sigma_0}{20m_e^2}
  \left[(E_\nu-\mu_\nu)^2+(\pi T)^2\right]\sqrt{\frac{\mu_\nu}{p_{F_e}}x_e} & \nonumber\\
		&\times\left[\left(C_{V,e}^2+C_{A,e}^2\right)\left(10+x_e^2\right)+10C_{V,e}C_{A,e}x_e\right],
\end{align}
\noindent for $E_\nu \gg T$. Here, $\sigma_0=4G_F^2m_e^2\hbar^2/\pi c^2\simeq 1.74\times 10^{-44}$ 
cm$^2$ and $x_e={\rm min}\left(\mu_\nu,p_{F_e}\right)/{\rm max}\left(\mu_\nu,p_{F_e}\right)$. 

Approximate expressions for neutrino-electron scattering for nondegenerate neutrinos were found for 
different neutrino energy regimes \cite{Tubbs1975}
\begin{align}\label{eq:Iw1982_scatt_NDnu_high}
	\frac{1}{\lambda_s^{\mbox{\scriptsize{ND}}}}&=\left[C_{V,e}^2+C_{A,e}^2+C_{V,e}C_{A,e}\right]
  \frac{n_e\sigma_0}{6m_e^2}E_\nu p_{F_e}
\end{align}
\noindent for $E_\nu\gg p_{F_i}$, and
\begin{align}\label{eq:Iw1982_scatt_NDnu_low}	  
	\frac{1}{\lambda_s^{\mbox{\scriptsize{ND}}}}&=\left[C_{V,e}^2+C_{A,e}^2\right]    
	\frac{n_i\sigma_0}{40 m_e^2}\frac{E_\nu^3}{p_{F_e}},
\end{align}
\noindent for $E_\nu\ll p_{F_e}$.

Neglecting the quark masses, the same results of 
Eqs. (\ref{eq:Iw1982_scatt_Dnu})-(\ref{eq:Iw1982_scatt_NDnu_low}) are expected to be valid for 
neutrino-quark scattering, with appropriate coupling constants. It is important to note that in all 
of Eqs. (\ref{eq:Iw1982_abs_Dnu})-(\ref{eq:Iw1982_scatt_NDnu_low}) it is assumed that also electrons 
are completely degenerate. Tubbs \& Schramm show approximations for neutrino-electron scattering 
when both neutrinos and electrons are nondegenerate \cite{Tubbs1975}. 

In Fig. \ref{fig:18} we show a comparison between full numeric results and analytic approximations 
of Eqs. (\ref{eq:Iw1982_abs_Dnu}) and (\ref{eq:Iw1982_abs_NDnu}) for the process 
$\nu_e+d\rightarrow e^-+\nu_e$. In the case of degenerate neutrinos, we have divided the numeric 
results by the factor $1-f_0(E_\nu)$, already included in the analytic results. In the nondegenerate 
case, we have used $\alpha_c=0.1$ in Eq. (\ref{eq:Iw1982_abs_NDnu}) (with $\alpha_c=0$ in the 
thermodynamic potential of quark matter). Larger values for $\alpha_c$ may result in better 
agreement with the numeric results for $E_\nu\gg T$ for this particular combination of temperature 
and density, but the same value for the strong coupling constant will not always give the better 
results for higher temperatures or densities for a given bag constant. 

Using Eq. (\ref{eq:Iw1982_abs_NDnu}) requires then fixing $\alpha_c$ to an optimal value, whenever 
this is physically acceptable, so the approximation may give the best results in its domain of 
validity. Eq. (\ref{eq:Iw1982_abs_Dnu}), for absorption of degenerate neutrinos, on the other hand 
is shown to give good results only on the vicinity of $E_\nu=\mu_\nu$ and the results are better 
the higher the neutrino degeneracy is.

In the upper panels of Fig. \ref{fig:19} we compare the results given by 
Eqs. (\ref{eq:Iw1982_scatt_NDnu_high}) and  (\ref{eq:Iw1982_scatt_NDnu_low}) with the numeric 
results for the indicated scattering processes of nondegenerate neutrinos. It is clear that the 
range of energies $E_\nu\lesssim T$ is in general not covered by the approximations. For the case 
of neutrino-electron scattering, Eq. (\ref{eq:Iw1982_scatt_NDnu_high}) gives the better results for 
the whole range of energies, even though its domain of validity is $E_\nu\gg p_{F_e}$. Nonetheless,
it must be noticed that, in this case, results share only the same order of magnitude and, roughly, 
the same qualitative behavior.

In the lower panels of of Fig. \ref{fig:19}, we show how the results of 
Eq. (\ref{eq:Iw1982_scatt_Dnu}) compare to the exact results. As in the case of absorption, the 
results are better for $E_\nu$ near $\mu_\nu$, nonetheless, the approximate results are in good 
qualitative agreement and same order of magnitude as the exact results also for $E_\nu < \mu_\nu$.
In the panel of the middle, we show how the strange quark mass affects the scattering cross section.

Using the approximate mean free paths given by 
Eqs. (\ref{eq:Iw1982_abs_Dnu})-(\ref{eq:Iw1982_scatt_NDnu_low}), we calculate the neutrino 
Rosseland mean free path and compare the results against exact calculations. Since we considered
only the cases of either highly degenerate or completely nondegenerate neutrinos, we did not use any 
interpolating procedure to join the different regimes, such as the one suggested by \cite{Keil1995}. 
Also, it must be noted that, for scatterings of nondegenerate neutrinos, only 
Eq. (\ref{eq:Iw1982_scatt_NDnu_low}) was used. Even though an interpolating scheme to join 
smoothly Eqs. (\ref{eq:Iw1982_scatt_NDnu_high}) and (\ref{eq:Iw1982_scatt_NDnu_low}) can be 
proposed, this is not expected to improve the results too much, since this energy range is strongly 
cut-off by the factor $1/\left(1-f_0(E_\nu)\right)$.

The results shown
in Fig. \ref{fig:20} reflect the fact that the approximations are, in general, good only on a 
limited range of neutrino energies. As one would expect, the analytic approximations of Eqs. 
(\ref{eq:Iw1982_abs_Dnu})-(\ref{eq:Iw1982_scatt_NDnu_low}) imply in generally poor agreement with 
exact results of energy-averaged opacities. Notoriously, the results are better for degenerate 
neutrinos. This can be understood from the fact that the approximations to the mean free paths tend 
to be better for neutrino energies near $\mu_\nu$ and the integrands of Eq. (\ref{eq:Lambda_Ross}) 
are highly weighted in the same region. Nevertheless, even in this case, the results agree only by 
order of magnitude and they share roughly the same qualitative behavior.

%%%%%%%%%%%%%%%%%%%%%%%%%%%%%%%%%%%%%%%%%%%%%%%%%%%%%%%%%%%%%%%%%%%%%%%%%%%%%%%%%%%%%%%%%%%%%%%%%%%% SEC
% SUMMARY / CONCLUSIONS                                                                            % CONCLU.
%%%%%%%%%%%%%%%%%%%%%%%%%%%%%%%%%%%%%%%%%%%%%%%%%%%%%%%%%%%%%%%%%%%%%%%%%%%%%%%%%%%%%%%%%%%%%%%%%%%%

\section{Summary and conclusions}

We have analyzed in detail the neutrino mean free paths in quark matter.  All the  
electron-neutrino weak interaction processes listed in Table \ref{table:cconsts} were considered. 
We have shown that, when neutrinos are nondegenerate, neutrino-quark scattering dominates the 
scattering contribution to the total neutrino opacity with respect to neutrino-electron scattering. 
On the other hand, when neutrinos are degenerate, the different scattering processes contribute 
almost the same, given that in this case, quark matter has a larger abundance of electrons.

In general, neutrino absorption constitutes the largest contribution to the neutrino opacity. 
However, neutrino scattering can represent a considerable fraction, amounting to about $40\%$ in the 
highly leptonized scenario. When neutrinos are nondegenerate, scattering represents no less than 
$10\%$ of the total opacity, for temperatures in the range $5-60$MeV and baryon number densitiy in 
the range $1-12n_0$. 

We estimate that using $1/\lambda_s$, given in Eq. (\ref{eq:kappa_0}) instead of $\kappa_1$ given 
by Eq. (\ref{eq:lambda_scatt}) in the computation of the neutrino scattering mean free paths will 
represent an error of less than $10\%$ in the total neutrino opacity. This upper bound is associated
with the threshold of neutrino free-streaming, where $f_1$ is known to be proportional to $f_0$ and 
both $1/\lambda_s$ and $\kappa_1$ can be calculated and compared. In regions where neutrinos are 
closer to thermodynamic equilibrium, and $f_1$ is vanishingly small, this error is expected to be 
much smaller.

For baryon number densities in the range $1-12n_0$, the energy-averaged diffusion coefficients 
$D_n$, show a greater dependency on temperature than on density when $T$ varies in the range 
$5-60$MeV. When neutrinos are nondegenerate, all of $D_2$, $D_3$ and $D_4$ decrease with baryon 
density, while when neutrinos are degenerate they show different behaviors. In this case, $D_2$ 
decreases, while $D_3$ show no significant variation and $D_4$ increases with increasing density. 
We have also shown that the diffusion coefficients associated with antineutrinos are larger than 
those of neutrinos when the latter are nondegenerate. When neutrinos are degenerate, on the other 
hand, the diffusion coefficients of neutrinos are greater than those of antineutrinos.

\begin{figure}[t]
\centering
\includegraphics[scale=0.8]{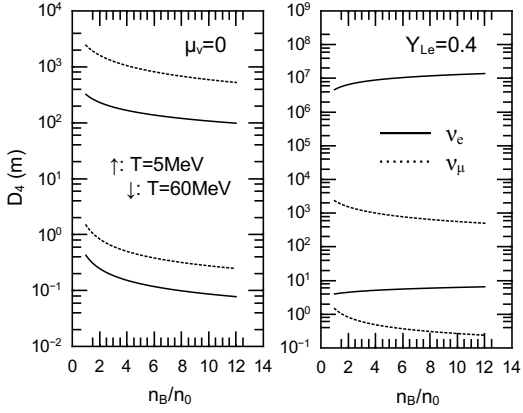}
\caption{\label{fig:21} Diffusion coefficient $D_4$ for electron neutrinos (full lines) and muon
neutrinos (dashed lines). The topmost pairs of curves correspond to $T=5$MeV and those at the bottom 
correspond to $T=60$MeV. Note that $\mu_\nu=0$
and $Y_{Le}=0.4$ correspond to electron neutrinos. Muon neutrinos always have $\mu_{\nu_\mu}=0$ here.}
\end{figure}

Since we have not considered the presence of muons and taus in quark matter, the corresponding 
curves associated to the diffusion of muon and tau neutrinos have been left
outside of the present work. Nevertheless, the following general remarks are worth addressing. 
Considering muon and tau neutrinos present in quark matter only in the form of thermally produced 
pairs, this type of neutrinos can take part only in the neutral-current scattering processes listed 
in Table \ref{table:cconsts}, given that the corresponding charged-current processes are kinematically
suppressed. It follows then that the total muon and tau neutrino mean free paths are 
generally increased in comparison with the total electron neutrino mean free paths. In particular, 
considering that the scattering opacity is the same for both muon and tau neutrinos, their 
contribution to the combined diffusion coefficient $D_4=D_{4}^{\nu_e}+D_4^{\bar{\nu}_e}+4D_4^{\nu_\mu}$
makes explicit their role in the energy transport in neutron stars, mainly in the regime of nondegenerate
electron neutrinos, as we show in Fig. \ref{fig:21}.

The influence of free parameters of the MIT bag model on the neutrino mean free paths have been 
analyzed. It is verified that increases in $B$, and $\alpha_c$ lead to in increases in the total 
neutrino opacity, while a higher $m_s$ decreses it. In general, when neutrinos are nondegenerate, 
increasing $B$ from $60$MeV/fm$^3$ to $200$MeV/fm$^3$ or increasing $m_s$ from $0$ to $150$MeV will 
affect more the neutrino opacity than does increasing $\alpha_c$ from $0$ to $0.6$ in the 
approximation of massless quarks. The same occurs when neutrinos are degenerate, with the difference 
that the effects are less prominent. 

It should be noted that we have not considered here how many-body interactions may alter the neutrino 
cross-sections through correlations or collective behavior in general. In specific, we have neglected 
the possibility of quark pairing and color superconductivity and how the neutrino mean free paths are 
altered by it. In fact, at asymptotically high densities it has been shown from first principles that 
color superconductivity occurs in QCD. However, in the density regime relevant for neutron star physics 
such weak-coupling calculations are unreliable and most conclusions are obtained from phenomenological 
models that are claimed to capture the essential physics of QCD. As a consequence, the actual occurrence 
of color superconductivity inside neutron stars is still an open question. Nevertheless, it is important 
to examine the impact of color superconductivity on the neutrino mean free paths and in fact this has 
been addressed in several works. For more details on the subject, we refer the reader to Refs. 
\cite{Carter2000,Reddy2003,Kundu2004,Schafer2004,Jaikumar2006,Pal2011}.

For last, neutrino-quark scattering share remarkable similarities with neutrino-electron scattering 
in what comes to neutrino/matter energy transfer and mean scattering angles. The general scenario 
is: low-energetic neutrinos tend to gain large amounts of energy and to scatter by larger angles. 
On the other hand, high-energetic neutrinos will lose on average half their initial energy and
scatter almost isotropically.

%%%%%%%%%%%%%%%%%%%%%%%%%%%%%%%%%%%%%%%%%%%%%%%%%%%%%%%%%%%%%%%%%%%%%%%%%%%%%%%%%%%%%%%%%%%%%%%%%%%% SEC
% Acknowledgments                                                                                  % ACKNOW.
%%%%%%%%%%%%%%%%%%%%%%%%%%%%%%%%%%%%%%%%%%%%%%%%%%%%%%%%%%%%%%%%%%%%%%%%%%%%%%%%%%%%%%%%%%%%%%%%%%%% 

\section*{Acknowledgments}
We would like to thank FAPESP and Universidade Federal do ABC for the financial support.

%%%%%%%%%%%%%%%%%%%%%%%%%%%%%%%%%%%%%%%%%%%%%%%%%%%%%%%%%%%%%%%%%%%%%%%%%%%%%%%%%%%%%%%%%%%%%%%%%%%% APPENDIX
% APPENDIX                                                                                         %
%%%%%%%%%%%%%%%%%%%%%%%%%%%%%%%%%%%%%%%%%%%%%%%%%%%%%%%%%%%%%%%%%%%%%%%%%%%%%%%%%%%%%%%%%%%%%%%%%%%% 

\appendix

%%%%%%%%%%%%%%%%%%%%%%%%%%%%%%%%%%%%%%%%%%%%%%%%%%%%%%%%%%%%%%%%%%%%%%%%%%%%%%%%%%%%%%%%%%%%%%%%%%%% APPENDIX
% APPENDIX A: LAMBDA                                                                               % A: LAMBDA
%%%%%%%%%%%%%%%%%%%%%%%%%%%%%%%%%%%%%%%%%%%%%%%%%%%%%%%%%%%%%%%%%%%%%%%%%%%%%%%%%%%%%%%%%%%%%%%%%%%% 

\section{Neutrino inverse mean free path}\label{sec:ap_mfp}
With $W_{fi}$ given in Eq. (\ref{eq:Wfi}), Eq. (\ref{eq:CollisionInt}) can be cast into several 
different forms for integration. To calculate the absorption mean free paths and the scattering mean 
free paths as in Eq. (\ref{eq:lambda_scatt}) it is convenient to use the total momentum as an 
integration variable. With that, all angular integrations can be performed and it only remains a 
double integration do be performed numerically 
\cite{Shalitin1978,Iwamoto1982,Gupta1995,Steiner2001}:

\begin{align}\label{eq:ap_mpf_lambda}
  \frac{1}{\lambda}=&\frac{gG_F^2}{32\pi^5} E_\nu^{-2}\int_{m_2}^\infty dE_2\int_{m_3}^\infty dE_3 
  E_3	\mathcal{S}\left[(C_V+C_A)^2I_a\right.\nonumber\\
	&\left.+(C_V-C_A)^2I_b-(C_V^2-C_A^2)I_c\right],
\end{align}
\noindent where $\mathcal{S}=f_0(E_2)[1-f_0(E_3)][1-f_0(E_1+E_2-E_3)]$ and the integrals $I_a$, 
$I_b$ and $I_c$ are given by
\begin{align}
    I_a=&\frac{\pi^2}{15}\left[3\left(P_{\mbox{\scriptsize{max}}}^5
    -P_{\mbox{\scriptsize{min}}}^5\right)\right.\nonumber\\
		&\quad\quad-10(k^{\nu 2}_++k^{34}_+)\left(P_{\mbox{\scriptsize{max}}}^3
    -P_{\mbox{\scriptsize{min}}}^3\right)\nonumber\\
		&\quad\quad\left.+60k^{\nu 2}_+k^{34}_+\left(P_{\mbox{\scriptsize{max}}}
    -P_{\mbox{\scriptsize{min}}}\right)\right],\\
		\nonumber\\
  	I_b=&I_a(E_2 \leftrightarrow -E_3),\\
		\nonumber\\
  	I_c=&\frac{2\pi^2}{3}m_2m_3\left[\left(Q_{\mbox{\scriptsize{max}}}^3
    -Q_{\mbox{\scriptsize{min}}}^3\right)\right.\nonumber\\
		&\quad\quad\left. +6k^{\nu 3}_-\left(Q_{\mbox{\scriptsize{max}}}
    -Q_{\mbox{\scriptsize{min}}}\right)\right],
\end{align}
\noindent with
\begin{align}
	&k_{\pm}^{ij}=E_iE_j\pm 0.5\left(p_i^2+p_j^2\right),\\
	&P_{\mbox{\scriptsize{min}}}=\mbox{max}\left(|E_\nu-p_2|,|p_3-p_4|\right),\\
	&P_{\mbox{\scriptsize{max}}}=\mbox{min}\left(E_\nu+p_2,p_3+p_4\right),\\
  &Q_{\mbox{\scriptsize{min}}}=|E_\nu-p_4|,\\
	&Q_{\mbox{\scriptsize{max}}}=\mbox{min}\left(E_\nu+p_4,p_2+p_3\right).
\end{align}
\noindent Here, $p_4=\sqrt{E_4^2-m_4^2}$ and the particle labels follow the convention of 
Table \ref{table:cconsts}. 

%%%%%%%%%%%%%%%%%%%%%%%%%%%%%%%%%%%%%%%%%%%%%%%%%%%%%%%%%%%%%%%%%%%%%%%%%%%%%%%%%%%%%%%%%%%%%%%%%%%% APPENDIX
% APPENDIX B: LAMBDA                                                                               % B: SCATT. KERNELS
%%%%%%%%%%%%%%%%%%%%%%%%%%%%%%%%%%%%%%%%%%%%%%%%%%%%%%%%%%%%%%%%%%%%%%%%%%%%%%%%%%%%%%%%%%%%%%%%%%%% 

\section{Scattering Kernels}\label{sec:scatt_kernel}
We define the neutrino scattering kernels $R^{out}$ as 
\begin{align}\label{eq:Rout_def}
  R^{out}&=g\int\frac{d^3p_2}{(2\pi)^3}\int\frac{d^3p_3}{(2\pi)^3}f(E_2)[1-f(E_3)]W_{fi},
\end{align}
\noindent where $g$ denotes the total phase-space degeneracy factor and the integrations involve 
only the target particles. With $W_{fi}$ given in Eq. (\ref{eq:Wfi}), all integrations except one 
can be performed exactly. The results can be found in references \cite{Yueh1976,Yueh1976a} (with 
the remark that our definition of $R^{out}$ does not carry the overall $E_\nu^{\prime 2}$):

\begin{align}\label{eq:Rout_ap}
	R^{out}=&\int_{E_{min}}^\infty dE_2 \mathcal{S}^\prime (E_\nu+E_2-E_\nu^\prime)
	[(C_V+C_A)^2h_a\nonumber\\
	&+(C_V-C_A)^2h_b-(C_V^2-C_A^2)h_c],
\end{align}
\noindent where $\mathcal{S}^\prime= f(E_2)[1-f(E_\nu+E_2-E_\nu^\prime)]$ and the integrals $h_a$, 
$h_c$ and $h_c$ are given by
\begin{align}
  h_a&=\frac{gG_F^2}{(2\pi)E_\nu E_\nu^\prime}\left[AE_2^2+BE_2+C\right],\\
	h_b&=h_a(E_\nu\leftrightarrow -E_\nu^\prime),\\
	h_c&=\frac{gG_F^2}{(2\pi)E_\nu E_\nu^\prime}
  \frac{m_2^2E_\nu E_\nu^\prime(1-\cos\theta)}{\alpha^{1/2}},
\end{align}
\noindent where $\alpha=E_\nu^2+E_\nu^{\prime 2}-2E_\nu E_\nu^\prime\cos\theta$, $\theta$ is the 
scattering angle. The quantities $A$, $B$ and $C$ are given by
\begin{align}
  A=&\frac{\gamma^2}{\alpha^{5/2}}\left[E_\nu ^2+E_\nu^{\prime 2}+(3+\cos\theta)E_\nu 
    E_\nu^\prime\right],\\
		\nonumber\\
	B=&\frac{\gamma^2}{\alpha^{5/2}}E_\nu \left[2E_\nu ^2+(3-\cos\theta)E_\nu 
    E_\nu^\prime\right.\nonumber\\
	  &\qquad\left.-(1+3\cos\theta)E_\nu^{\prime 2}\right],\\
		\nonumber\\
	C=&\frac{\gamma^2}{\alpha^{5/2}}E_\nu ^2\left[(E_\nu -E_\nu^\prime\cos\theta)^2
    -\frac{E_\nu^{\prime 2}\sin^2\theta}{2}\right.\nonumber\\
	  &\qquad\left.-\frac{m_2^2(1+\cos\theta)}{2E_\nu ^2(1-\cos\theta)}\alpha\right],
\end{align}

\noindent where we have also defined $\gamma=E_\nu E_\nu^\prime(1-\cos\theta)$. The lower 
integration limit in Eq. (\ref{eq:Rout_ap}), is given by
\begin{align}
  E_{min}=\frac{E_\nu^\prime-E_\nu}{2}+\frac{\alpha^{1/2}}{2}\sqrt{1+\frac{2m_2^2}{\gamma}}.
\end{align}

%%%%%%%%%%%%%%%%%%%%%%%%%%%%%%%%%%%%%%%%%%%%%%%%%%%%%%%%%%%%%%%%%%%%%%%%%%%%%%%%%%%%%%%%%%%%%%%%%%%% APPENDIX
% APPENDIX C: ETRANSFER / SANGLE                                                                   % C: ET / SANGLE
%%%%%%%%%%%%%%%%%%%%%%%%%%%%%%%%%%%%%%%%%%%%%%%%%%%%%%%%%%%%%%%%%%%%%%%%%%%%%%%%%%%%%%%%%%%%%%%%%%%%

\section{Mean energy transfer per scattering and mean scattering angle}\label{Ap:etransfer}
From Fermi's golden rule, the neutrino interaction cross section can be written in terms of the 
squared matrix element, summed over final spins and averaged over the initial spins
\begin{multline}
  \sigma=\frac{(2\pi)^{-6}}{4(p_\nu\cdot p_2)}\int \frac{d^3p_3}{2E_3}\int\frac{d^3p_4}{2E_4}
  \langle\mathcal{M}\rangle^2
	\\
	\times(2\pi)^4\delta(p_\nu+p_2-p_3-p_4),
\end{multline}
\noindent for interactions happening on the vacuum. In thermodynamic systems, the particles involved 
in such scattering reactions have specific energy distributions and we must average the cross 
sections over their distribution functions and the availability of the initial and final states. 
In this case, we write 
\begin{align}\label{eq:sigbar}
	\bar{\sigma}=&(2\pi)^{-5}g\int \frac{d^3p_2}{2E_2}\int \frac{d^3p_3}{2E_3}\int \frac{d^3p_4}{2E_4}
	f_0(E_2)[1-f_0(E_3)]\nonumber\\
	&\times[1-f_0(E_4)]\frac{\langle\mathcal{M}\rangle^2}{4\left(p_\nu\cdot p_2\right)}
	\delta(p_\nu+p_2-p_3-p_4).
\end{align}

We define the mean energy transfer per collision between neutrinos 
and matter and the mean scattering angle as the averages
\begin{flalign}\label{eq:etransf}
	\delta E_\nu =&\frac{g(2\pi)^{-3}}{\bar{\sigma}}\int \frac{d^3p_2}{2E_2}\int dE_\nu^\prime 
  \mathcal{S}
	\left(E_\nu-E_\nu^\prime\right)\left(\frac{\partial\sigma}{\partial E_\nu^\prime}\right),&
\end{flalign}
\noindent and
\begin{flalign}\label{eq:sangle}
	\langle 1-\cos\theta\rangle =&\frac{g(2\pi)^{-3}}{\bar{\sigma}}\int \frac{d^3p_2}{2E_2}\int 
  dE_\nu^\prime \mathcal{S}&\nonumber\\
	&\times\int d\cos\theta(1-\cos\theta)\left(\frac{\partial^2\sigma}{\partial E_\nu^\prime
    \partial \cos\theta}\right),&
\end{flalign}
\noindent where we have defined
\begin{align}
  \mathcal{S}=f_0(E_2)[1-f_0(E_\nu+E_2-E_\nu^\prime)][1-f_0(E_\nu^\prime)].
\end{align}

Note that we have added the blocking factor $1-f_0(E_\nu^\prime)$, to account for neutrino 
degeneracy, in contrast to the results of \cite{Tubbs1975}. Also, our definition of $\bar{\sigma}$ 
is only a shorthand for the quantity defined in Eq. (\ref{eq:sigbar}) and not a properly averaged 
statistical cross section as the one found in the same reference. For convenience, we have added a 
minus overall sign in Eq. (\ref{eq:etransf}) for the mere fact that we want negative values to 
represent energy loss.

Substituting $\langle\mathcal{M}\rangle^2$ on Eqs. (\ref{eq:etransf}) and (\ref{eq:sangle}), we 
obtain Eqs. of the form
\begin{align}
  \delta E_\nu =\frac{1}{\bar{\sigma}}&\left[(C_V+C_A)^2W_1+(C_V-C_A)^2W_2\right.\nonumber\\
	&\left.+(C_A^2-C_V^2)W_3\right]
\end{align}
\noindent and
\begin{align}
  \langle 1-\cos\theta\rangle =\frac{1}{\bar{\sigma}}&\left[(C_V+C_A)^2U_1+(C_V-C_A)^2U_2
  \right.\nonumber\\ 
	&\left.+(C_A^2-C_V^2)U_3\right]
\end{align}
\noindent with the $Ws$ and $Us$ given by Eqs. (21a)-(22c) of \cite{Tubbs1975} with the inclusion of 
$[1-f_0(\omega+\epsilon-\epsilon^\prime)]$ in all $\epsilon^\prime$ integrations of the cited 
reference.

We would like to point slightly different definitions for $\delta E_\nu$ and 
$\langle1-\cos\theta\rangle$ that may be derived directly from Eq. (\ref{eq:lambda_scatt}):
\begin{align}\label{eq:etransf_alt}
  \delta E_\nu=-\lambda_s\int_0^\infty dE_\nu^\prime 
	(E_\nu-E_\nu^\prime)\left[\frac{\partial(1/\lambda_s)}{\partial E_\nu^\prime}\right],
\end{align}
\noindent and
\begin{align}\label{eq:scattang_alt1}
  \langle1-\cos\theta\rangle =&\lambda_s\int dE_\nu^\prime\int d\cos\theta
	\nonumber\\
	&\times(1-\cos\theta)\left[\frac{\partial^2(1/\lambda_s)}
  {\partial E_\nu^\prime\partial\cos\theta}\right].
\end{align}

\noindent Eq. (\ref{eq:scattang_alt1}) can then be written in terms of the Legendre moments of the 
scattering kernel $R^{out}$:
\begin{flalign}\label{eq:scattang_alt}
  &\langle1-\cos\theta\rangle =1-\frac{\lambda_s}{(2\pi)^2}\int dE_\nu^\prime E_\nu^{\prime 2}
  \left[1-f_0(E_\nu^\prime)\right]R^{out}_1.&
\end{flalign}

\noindent
Even though Eqs. (\ref{eq:etransf_alt}) and (\ref{eq:scattang_alt}) differ from 
Eqs. (\ref{eq:etransf}) and (\ref{eq:sangle}) by the lack of the factor $(p_\nu\cdot p_2)^{-1}$ on 
the integrands (coming from the definition of cross section), they give fairly similar results and 
they can be more easily evaluated when $R^{out}$ is available.

%%%%%%%%%%%%%%%%%%%%%%%%%%%%%%%%%%%%%%%%%%%%%%%%%%%%%%%%%%%%%%%%%%%%%%%%%%%%%%%%%%%%%%%%%%%%%%%%%%%% APPENDIX
% APPENDIX D: ETRANSFER / SANGLE                                                                   % D: ANA. Dn
%%%%%%%%%%%%%%%%%%%%%%%%%%%%%%%%%%%%%%%%%%%%%%%%%%%%%%%%%%%%%%%%%%%%%%%%%%%%%%%%%%%%%%%%%%%%%%%%%%%%

\section{Analytic approximation to the diffusion $D_n$ coefficients for degenerate neutrinos}
\label{Ap:Dn}
In the case of highly degenerate neutrinos, the analytic approximations to the neutrino mean
free paths with respect to both the neutral and charged currents, 
Eqs. (\ref{eq:Iw1982_abs_Dnu})-(\ref{eq:Iw1982_scatt_Dnu}), can be combined in the form
\begin{align}\label{eq:analytic_DNu_lambda_g}
  \frac{1}{\lambda_T}&=\left[\left(x-\eta_\nu\right)^2+\pi^2\right]\frac{1}{g(\eta_\nu)},
\end{align}
\noindent where
\begin{align}
  \frac{1}{g(\eta_\nu)}=\frac{1}{\eta_\nu^2}\left(a + b\eta_\nu^3+ c\eta_\nu^4+d\eta_\nu^5\right)
\end{align}
\noindent with the coefficients given by
%\begin{widetext}
\begin{align}
  a&=\left[\frac{ G_F^2 T^5}{3\pi^3 \left(c\hbar\right)^7}\right] 12\cos^2\theta_c  
      \eta_u^2\eta_e^3\left[1+\frac{1}{2}\frac{\eta_e}{\eta_u}\right. \nonumber\\
			&\qquad\left.+\frac{1}{10}\left(\frac{\eta_e}{\eta_u}\right)^2\right],\\
	{}
	b&=\left[\frac{ G_F^2 T^5}{3\pi^3 \left(c\hbar\right)^7}\right]\sum_{i=u,d,s,e}g_i
     \left(C_{V,i}^2+C_{A,i}^2\right)\eta_i^2,\\
	{}
	c&=\left[\frac{ G_F^2 T^5}{3\pi^3 \left(c\hbar\right)^7}\right]\sum_{i=u,d,s,e}g_iC_{V,i}C_{A,i}
     \eta_i,\\
	{}
	d&=\left[\frac{ G_F^2 T^5}{3\pi^3 \left(c\hbar\right)^7}\right]\frac{1}{10}\sum_{i=u,d,s,e}g_i
    \left(C_{V,i}^2+C_{A,i}^2\right).
\end{align}
%\end{widetext}

\noindent In these equations, $\eta_i=p_{F_i}/T$. Coefficients $b$ to $d$ are related only to the 
scattering reactions. Coefficient $a$ has a counterpart as mentioned after 
Eq. (\ref{eq:Iw1982_abs_Dnu}), however, $|p_F(u)-p_F(e)|\geq |p_F(d)-\mu_\nu|$, tends to be the 
case when neutrinos are highly degenerate. Note also that we used $x_i=\mu_\nu/p_{F_i}$ in 
Eq. (\ref{eq:Iw1982_scatt_Dnu}).

In this regime, the diffusion coefficients $D_n$ read
\begin{align}\label{eq:Dn_analytic_first}
  D_n&=g(\eta_\nu)
 \int_0^\infty dx x^n \left\{\frac{f_0(E_\nu)\left[1-f_0(E_\nu)\right]}
  {\left(x-\eta_\nu\right)^2+\pi^2}\right\}.
\end{align}
\noindent The expression between curly braces can be approximated by
\begin{align}
  \frac{f_0(E_\nu)\left[1-f_0(E_\nu)\right]}{\left(x-\eta_\nu\right)^2+\pi^2}\simeq
	\frac{e^{-\left(x-\eta_\nu\right)^2/4}}{4\pi^2},
\end{align}
\noindent and so Eq. (\ref{eq:Dn_analytic_first}) can be solved by Gamma functions. One then
arrives at
\begin{multline}
  D_n=g(\eta_\nu)\frac{1}{4\pi^2}\sum_{k=0}^n\binom{n}{k}4^{k/2}\left[1+(-1)^k\right]\\
	\times\Gamma\left(\frac{k+1}{2}\right)\eta_\nu^{n-k}.
\end{multline}
\noindent Explicitly:
\begin{align}
  D_2&=\frac{1}{2\pi^{3/2}}\left(\frac{2\eta_\nu^2+\eta_\nu^4}{a+b\eta_\nu^3+c\eta_\nu^4
        +d\eta_\nu^5}\right)\label{eq:D2_approx},\\
  D_3&=\frac{1}{2\pi^{3/2}}\left(\frac{6\eta_\nu^3+\eta_\nu^5}{a+b\eta_\nu^3+c\eta_\nu^4
        +d\eta_\nu^5}\right)\label{eq:D3_approx},\\
  D_4&=\frac{1}{2\pi^{3/2}}\left(\frac{12\eta_\nu^2+12\eta_\nu^4+\eta_\nu^6}{a+b\eta_\nu^3
        +c\eta_\nu^4+d\eta_\nu^5}\right)\label{eq:D4_approx}.	
\end{align}

To perform a quick analysis, we assume, for simplicity, that 
$\eta_u\simeq \eta_d\simeq\eta_e\simeq\eta_\nu$ (which is, in fact, a reasonable assumption when 
neutrinos are highly degenerate). With this, the denominator on 
Eqs. (\ref{eq:D2_approx})-(\ref{eq:D4_approx}) reduces to
\begin{align}
  a+b\eta_\nu^3+c\eta_\nu^4+d\eta_\nu^5\simeq \left[\frac{ G_F^2 T^5}{3\pi^3 
  \left(c\hbar\right)^7}\right] 19.78 \eta_\nu^5.
\end{align}
\noindent It is clear then that coefficients $D_2$, $D_3$ and $D_4$ have different behavior with 
respect to $\eta_\nu$ and, in particular, $D_4$ will increase with increasing neutrino degeneracy.

%\bibliography{opac}


\end{document}